\documentclass[twocolumn, linenumbers]{aa}  
\usepackage{adjustbox}
\usepackage{enumitem}
\usepackage{hyperref}
\usepackage{graphicx}
\usepackage{subcaption}
\usepackage{txfonts}
\usepackage{soul}
\usepackage[dvipsnames]{xcolor}
\usepackage{colortbl}
\usepackage{ulem}
\usepackage[labelfont=bf]{caption}

\newcommand{\Fermi}{\textit{Fermi}\xspace}
\newcommand{\TXS}{TXS~0506$+$056\xspace}

\catcode`\@=11 \input numdef.sty \catcode`\@=12

\num\newcommand{\J2115}{4FGL~J2115.9$-$0113\xspace}
\num\newcommand{\J1258}{4FGL~J1258.7$-$0452\xspace}
\num\newcommand{\J1438}{4FGL~J1438.6$+$1205\xspace}
\num\newcommand{\J1342}{4FGL~J1342.7$+$0505\xspace}
\num\newcommand{\J0604}{4FGL~J0604.9$-$0000\xspace}
\num\newcommand{\J0127}{4FGL~J0127.3$-$0148\xspace}

\newcommand{\paperPHone}{\textsc{Paper~PI}} 
\newcommand{\paperPHthree}{\textsc{Paper~PIII}}

\newcommand{\accretion}{L_{\rm BLR} / L_{\rm Edd}}
\newcommand{\gratio}{L_{\gamma} / L_{\rm Edd}}
\newcommand{\Lg}{L_{\gamma}}
\newcommand{\Pradio}{P_{1.4\,{\rm GHz}}}
\newcommand{\Msun}{M_{\odot}}
\newcommand{\mbh}{M_{\rm BH}}

\newcommand{\WHz}{{\rm W}\cdot{\rm Hz}^{-1}}

\begin{document}

\title{Lepto-hadronic modeling of blazars associated with well-reconstructed high-energy neutrino events}

\author{L. Pfeiffer\inst{1}\corrauth{leonard.pfeiffer@uni-wuerzburg.de}
\and M. Boughelilba\inst{2}\corrauth{margot.boughelilba@gmail.com}
\and S. Garrappa\inst{3}\email{simone.garrappa@weizmann.ac.il} 
\and S. Buson\inst{1,2}\email{sara.buson@gmail.com}  
\and M. Lincetto\inst{2}\email{lincetto@cppm.in2p3.fr} 
\and A. Bremer\inst{1}\email{annette.bremer@stud-mail.uni-wuerzburg.de}}

\institute{ Julius-Maximilians-Universit{\"a}t W{\"u}rzburg, Fakult{\"a}t f{\"u}r Physik und Astronomie, Institut f{\"u}r Theoretische Physik und Astrophysik, Lehrstuhl f{\"u}r Astronomie, Emil-Fischer-Str. 31, D-97074 W{\"u}rzburg, Germany \and Deutsches Elektronen-Synchrotron DESY, Platanenallee 6, 15738 Zeuthen, Germany \and Department of Particle Physics and Astrophysics, Weizmann Institute of Science, 76100 Rehovot, Israel}

  \abstract
   {The association of active galactic nuclei (AGN) with high-energy neutrinos has been investigated in both population studies and individual source analyzes. However, whether individual associations are driven by physical processes or arise from chance coincidences remains an open question.}
   {A sample of blazars, AGN with a relativistic jets closely aligned with the line of sight, spatially consistent with IceCube high-energy events, provides a natural framework to address this question. By combining multifrequency observations with physically motivated emission models, it is possible to assess whether these sources can plausibly account for the detected neutrinos.}
   {We study six blazars with single gamma-ray counterparts to well-reconstructed IceCAT-1 high-energy events, accounting for the improved localization expected from IceCAT-2, and compile their multifrequency spectral energy distributions. We model these sources within a lepto-hadronic framework, exploring their parameter space in order to reproduce the observed emission while preferentially selecting solutions that maximize neutrino production.}
   {We find that the spectral energy distributions can be reproduced with compact emission regions close to the broad-line region with moderately high Doppler factors. The overall emission is generally dominated by leptonic processes with subdominant contributions from hadronic cascades, except for one source where the hadronic components become dominant in gamma-rays. For four of six sources, the energy density of the relativistic particles is close to equipartition with the magnetic field. The predicted neutrino detection rates reach up to $\sim5\times10^{-2}$ per year, in most cases compatible with the observation of one IceCube high-energy event when accounting for Poisson statistics.}
   {We conclude that, while the sources in the sample are not expected to significantly contribute to the diffuse neutrino flux, their properties are consistent with being possible counterparts to individual neutrino events. This is driven by a significant contribution of external photon fields to the hadronic interaction rate, even for sources where these fields have previously only been considered as non significant.}


   \maketitle
   \nolinenumbers
\section{Introduction}
The IceCube Neutrino Observatory, which is currently the most sensitive instrument for astrophysical neutrinos, has established the existence of a diffuse flux of astrophysical neutrinos at high energies \cite{Aartsen_instrument}. However, the origin of the dominant contributors to this flux remains largely unknown. Among the proposed source classes, active galactic nuclei with relativistic jets pointing in the direction of earth, so called blazars, represent particularly compelling candidate neutrino sources due to their large non-thermal luminosities, and the potential for efficient hadronic acceleration \citep[e.g.][]{mannheim1993protonblazar,MANNHEIM1995295, Atoyan:2003,PhysRevD.88.047301,DERMER201429, PhysRevD.90.023007, 10.1093/mnras/stv179, Keivani_2018, 2013ApJ...768...54B, 10.1093/mnrasl/sly210, PhysRevLett.126.191101,refId0}. In such environments, high-energy neutrinos may be produced via pion production in p-$\gamma$ or pp interactions. Observational support for this scenario was strengthened by the association of the flaring $\gamma$-ray blazar \TXS with the high-energy neutrino event IC-170922A \citep{Icecube_TXS_flaring:2018}, as well as time-integrated studies of the IceCube data, that suggest statistically significant anisotropies in the spatial distribution of IceCube event sky-maps to be consistent with a sample of blazars, including \TXS \citep{Buson:2022,Buson_erratum:2022,Buson:2023,Bellenghi:2023}.
Systematic follow-up analyses of high-energy IceCube alerts, such as presented by \citet{Garrappa:2024}, have shown that only a small number of events positionally coincident with known $\gamma-$ray blazars could be identified as potential multi-messenger associations. However, the post-trial chance probabilities are generally consistent with background expectations \citep[e.g.][]{IC_north_hard_spectrum:2016,IceCube7y:2017,IceCube2017_2LAC, Icecube_TXS_flaring:2018,IceCube10y:2020,IceCube_10y_reprocessed:2022,IceCube_AGNcores:2021}. Population-based studies further indicate that while blazars are unlikely to dominate the diffuse astrophysical neutrino flux, they may populate a sub-dominant yet potentially identifiable component \citep{Aartsen_20172LAC, Abbasi_2023radio}. To date, the identification of sources similar to \TXS has relied on single-event spatial coincidences without detailed physical assessment of individual source candidates in a multifrequency context.\\
Since the latest publication of $\gamma-$ray follow-up studies \citep{Garrappa:2019,Garrappa:2024}, new high-energy neutrinos have been observed. Furthermore also the Fourth \Fermi-LAT Gamma-ray Source Catalog \citep[4FGL,][]{ballet2024} and  the Fourth Catalog of Active Galactic Nuclei \citep[4LAC,][]{4LAC_DR3:2022} were updated and refined. In parallel, the IceCube Collaboration has developed an improved reconstruction pipeline, which was applied for alerts starting in late 2024 \citep{IceCAT-2:2025} and will be applied also for older alerts in the forthcoming IceCube Event Catalog of Alert Tracks version 2 (IceCAT-2). Such developments highlight the need of reassessing previously reported neutrino$-$source associations. 

Rather than maximizing the number of spatial coincidences, this work focuses on a selected sample of well-reconstructed high-energy events with unique blazar counterparts, explicitly accounting for the changes in event localization expected for the IceCAT-2 reconstruction. This strategy yields a counterpart candidate sample of high-purity particularly suited for physical interpretation, predictive modeling and energetic feasibility tests in the context of its overall multifrequency behavior. For each selected source, we assess whether neutrino production is energetically viable and derive source specific neutrino flux expectations. For this work we focus on a sample of six blazars derived from criteria introduced in Section \ref{sec: sample}. In Section \ref{sec: observational data} we describe their multifrequency data collection which is used to derive physical properties and build broad-band spectral energy distributions (SED) in Section \ref{sec: mwl}. The modeling procedure described in Section \ref{sec: mwl} is applied to the data in Section \ref{sec: result} and discussed in Section \ref{sec: discussion}.

\section{Sample selection}\label{sec: sample}
\subsection{The IceCube high-energy neutrino alerts}
In this work, we focus on high-energy neutrino events observed by IceCube, with typical reconstructed energies $\geq$ 100 TeV. At these energies, the astrophysical component dominates over the atmospheric background, and the selected events are expected to have a substantial probability of astrophysical origin. Given the low detection rate ($\simeq$ 2 per month) and their importance for multi-messenger searches, a dedicated realtime alert stream was introduced in 2016 \citep{RealTimeStream1} to trigger prompt multifrequency follow-up observations by the astronomical community. Early alert classifications were based on event topology, separating events into extremely high energy (EHE) and high-energy starting events (HESE), each with an associated probability of astrophysical origin (“signalness”). Since 2019, alerts have been classified primarily by signalness into two classes, Gold and Bronze, with average signalness values of 50\% and 30\%, respectively.

The first catalog of these events is IceCAT-1 \citep{IceCAT-1:2023}, which includes both realtime-dispatched events and archival events detected prior to 2016 that satisfy, a posteriori, the realtime-stream selection. More recently, an updated release has been announced by the IceCube Collaboration (IceCAT-2; \citealt{IceCAT-2:2025}), which will also provide improved reconstructions for events included in IceCAT-1.

\subsection{Well-reconstructed neutrino events}
Here we use a sample of high-energy neutrino events distributed through the IceCube realtime alert streams since 2016, as well as archival events that satisfy a posteriori the realtime-stream selection and are included in the IceCAT-1 catalog. This collection is based on the sample used in \citealt{Garrappa:2024}, with the addition of more recent realtime events. In total, 394 neutrino events are included in the starting sample, covering more than 14 years of IceCube observations from 2011 May 14 (IC110514A) to 2025 July 8 (IC250708A). In this sample, for all the realtime events that have been included in the v4 of IceCAT-1, we use the most updated reconstruction reported in the catalog.\\
The 90\% containment regions of these events span a wide range of sky areas, from as small as 0.15~sq. deg to as large as 721~sq. deg. Therefore, before any catalog cross-matching, it is important to select a subsample of well-reconstructed alerts to reduce the chance of random coincidences in multi-messenger associations. Following \citealt{Garrappa:2024}, we consider the observed median extension of the 90\% containment regions of the Gold alerts sample (4.78 sq. deg) as a threshold for relatively well-reconstructed events.  
After selection, the final sample of relatively well-reconstructed alerts contains 149 events. Among those, 53 are classified as Gold, 65 as Bronze, 17 as EHE, and 14 as HESE.
\subsection{Single \Fermi-LAT blazar coincidences}\label{sec: gamma_neutrino_connection}
We crossmatch the sample of relatively well-reconstructed neutrinos with the fourth \Fermi-LAT catalog of Active Galactic Nuclei (4LAC-DR3, \citealt{4LAC_DR3:2022}) by selecting the gamma-ray sources within the neutrino contour. We find a total of 18 high-energy neutrinos from the sample of relatively well-reconstructed alerts which have a single gamma-ray blazar coincidence. These associations are listed in Table \ref{tab:coincidence_table}, including their most recent \Fermi-LAT catalog name to date (see Section \ref{sec: gamma-ray}), multifrequency candidate counterpart, neutrino event type and signalness ($\mathrm{p}_s$). Sources shaded in gray do not satisfy the selection criteria described in Section \ref{sec: rcross} and are excluded from the primary sample.
\begin{table*}[h!]
\centering
\caption{4LAC-DR3 sources positionally consistent with well-reconstructed realtime alerts from Section \ref{sec: gamma_neutrino_connection}.}
\resizebox{\textwidth}{!}{\begin{tabular}{lllllccc}
\hline
4FGL Name &  FL16Y Name&Counterpart & Event & Type & $\mathrm{p}_s$  & $r_{\rm cross}$ & Note\\ \hline
 J0127.3-0148& -& FBQS J0127-0151& IC170809A& Gold& 0.60&0.67 &\\
 \rowcolor{gray!15} J0244.7+1316& J0244.5+1316& GB6 J0244+1320& IC161103A& Bronze& 0.32&0.87 &(a)\\
 \rowcolor{gray!15} J0347.6-0133& J0347.5-0132& NVSS J034732-013218& IC190113A& Bronze& 0.39&0.71 &(b)\\
 \rowcolor{gray!15} J0420.3-3745 & J0420.3-3744& NVSS J042025-374443  & IC190504A & Bronze & 0.39  &0.49 &(b)\\
 \rowcolor{gray!15} J0436.2-0038 & J0436.3-0035& NVSS J043614-003637& IC180608A & Bronze &0.40  &0.47 &(b)\\
 \rowcolor{gray!15} J0509.4+0542 &  J0509.4+0541&\TXS & IC170922A  & EHE  & 0.51                 &0.12 &(c)\\
 \rowcolor{gray!15} J0515.9+0537& -& TXS 0513+054& IC160104A& Gold& 0.57&0.71 &(b)\\
 J0604.9-0000 & J0604.9-0000& GB6 J0604+0000 & IC130822A & Bronze & 0.30  &0.55 &\\
 \rowcolor{gray!15}J0658.6+0636& J0658.6+0637& NVSS J065844+063711& IC201114A& Gold& 0.56&1.28 &(a)\\
 \rowcolor{gray!15}J1204.8+0407& J1204.8+0402& MG1 J120448+0408& IC120916A& Gold& 0.44&1.05 &(a)\\
 J1258.7-0452 & J1258.7-0450& RBS 1194& IC150926A      & EHE    & 0.30 &0.28 &\\
 J1342.7+0505 & J1342.7+0503& 4C +05.57& IC210210A & Gold &0.65  &0.48 &\\
 J1438.6+1205& J1438.6+1204& RX J1438.3+1204& IC190413A& Bronze&0.29&0.44 &\\
 \rowcolor{gray!15}J1506.6+0813& J1506.7+0813& PMN J1506+0814& IC121115A& Bronze& 0.32&1.17 &(a)\\
 \rowcolor{gray!15} J1727.2+0644& J1725.8+0655& NVSS J172720+064123& IC250102A& Bronze& 0.26&0.95 &(a)\\
 \rowcolor{gray!15}J1744.2-0353& J1744.0-0351& PKS 1741-03& IC110930A& EHE& 0.43&1.18 &(a)\\
 \rowcolor{gray!15}J1916.7-1516& J1916.5-1518& PMN J1916-1519& IC131204A& EHE& 0.20&1.13 &(a)\\
 J2115.9-0113 & J2115.9-0110& NVSS J211603-010828 & IC250101A & Bronze &0.41  &0.22 &\\
\end{tabular}}
\begin{flushleft}
\textbf{Notes.} The source list is ordered by right ascension. Sources shaded in gray do not satisfy the selection criteria and are excluded from the primary sample for the following reasons:\\
\footnotesize 
(a) The source does not pass the crossmatch criterion defined in Section \ref{sec: rcross}.\\
(b) The source has no publicly available constraint on physical properties or redshift (Section \ref{sec: rcross} and \ref{subsec: optical spectra})\\
(a) The source is well studied in the literature.
\end{flushleft}
\label{tab:coincidence_table}
\end{table*}

\subsection{Toward IceCAT-2}\label{sec: rcross}
The IceCube Collaboration recently presented improvements to its event reconstruction pipeline, which will be reflected in the upcoming IceCAT-2 catalog \citep{IceCAT-2:2025}. They report systematic shifts and changes in the containment region size between the reconstructed positions in IceCAT-1 and those obtained with the updated reconstruction.
 We estimate a cross-matching distance that accounts for both the expected positional shift and the angular uncertainty of the event. Specifically, we compute $r_{\mathrm{cross}}=\sqrt{r_{\mathrm{shift},50\%}^2+r_{\mathrm{cont},90\%}^2}$,
where $r_{\mathrm{shift},50\%}$ represents the median positional shift (corresponding to the $50\%$ quantile of the IceCAT-1/IceCAT-2 angular separation from \cite{IceCAT-2:2025}), and $r_{\mathrm{cont},90\%}$ is the radius derived from the median $90\%$ containment area calculated for IceCAT-2. This expression combines the expected systematic displacement with the statistical localization uncertainty, providing a conservative estimate of the region in which the true counterpart is likely to remain after the reconstruction update.
For each event, we infer the angular distance between the best-fit positions of each of the 18 selected neutrinos and their nearby 4LAC-DR3 source. We prioritize associations when the separation is smaller than $r_{\mathrm{cross}}$.\\
After this additional selection, we remove eight neutrino-blazar coincidences that do not satisfy the predictions for association in IceCAT-2, reducing the sample to ten neutrino blazar candidates.\\
Moreover, we do an extensive search of archival multifrequency data for each source (see Sec. \ref{sec: observational data}). For the robustness of the modeling of the selected candidates, we exclude sources that lack both optical spectroscopic constraints and a measured redshift. Sources are retained if at least one of these two pieces of information is available. The final sample of sources which will be modeled consists of six neutrino blazar candidates listed in Table \ref{tab: properties from optical}.

\section{Observational data}
\label{sec: observational data}

The emission of blazars across the electromagnetic spectrum is generally explained by processes involving relativistic particles within the jet and surrounding environment. The resulting SEDs are commonly described by leptonic scenarios like synchrotron radiation from accelerated electrons and subsequent inverse Compton scattering of low-energy photons. This picture can be extended to invoke a mostly subdominant hadronic component, which accounts for relativistic protons interacting with radiation fields and ambient matter. Through production of secondary particles in this environment, charged pion decay provides a natural channel for high-energy neutrino production. For the primary sample of neutrino-associated blazars, we collected and analyzed broadband data taken from 2004 to 2025. In the following sections we will highlight the instruments and catalogs involved in the data collection while a summary of the applied standard data reduction methods is provided in e.g. \citet{lincettoplavinpfeiffer2025}.
\subsection{Radio to ultra-violet data}
Starting at radio frequencies, we collected the values of the power at $1.4\,{\rm GHz}$ from the National Radio Astronomy Observatory (NRAO) Very Large Array (VLA) Sky Survey \citep[NVSS Catalog,][]{NVSS_survey:1998} for the full sample of six blazars. This property was, then, used to trace the intrinsic power of their relativistic jets as further explained in Section \ref{subsec: optical spectra}. We furthermore used catalogs from \textit{Specfind} V3.0 \citep{SPECFIND} to gather additional radio flux measurements at multiple frequencies, which allowed us to extend the spectral coverage of the sources and later build more complete SEDs.
Also Infrared (IR) observations performed with the Wide-field Infrared Survey Explorer \cite[\textit{WISE};][]{WISE,WISEfilters} and the Near-Earth Object Wide-field Infrared Survey Explorer \cite[\textit{NEOWISE};][]{NEOWISE} are gathered from public archival data. While WISE observed magnitudes in the W1 ($3.4\,\mu\mathrm{m}$), W2 ($4.6\,\mu\mathrm{m}$), W3 ($11.6\,\mu\mathrm{m}$) and W4 ($22.1\,\mu\mathrm{m}$) band, NEOWISE continued the mission only with the W1 and W2 filter.
Public archival optical data are collected from the Panoramic Survey Telescope And Rapid Response System \cite[\textit{PAN-STARRS};][]{PANSTARRS1,PANSTARRS2, PANSTARRS3, PANSTARRS4,PANSTARRS5, PANSTARRS6} available in five optical to infrared filters (z,y,i,g,r), the Asteroid Terrestrial-impact Last Alert System \citep[ATLAS;][]{ATLAS1,ATLAS2,ATLAS3} with two optical filters (orange (o) and cyan(c)), the Catalina Real-Time Transient Survey \cite[CRTS;][]{CRTS}, the Zwicky Transient Facility \citep[ZTF;][]{ZTF} with three available filters (r,g,i) which was built on the proven heritage of the Palomar Transient Facility \citep[PTF;][]{PTF} used here with its R and g filters.
We furthermore include data from the \textit{Gaia} mission \citep{Gaia1,Gaia2,Gaia3} with its custom G, $\mathrm{G}_{RP}$ and $\mathrm{G}_{BP}$ filters, as well as observations from the All-Sky Automated Survey for Supernovae \citep[ASAS-SN;][]{ASAS1, ASAS2} which uses V and g filters. Observations in optical to ultra-violet wavelengths are obtained by reducing observations performed with the Ultraviolet/Optical-Telescope \citep[\textit{Swift-}UVOT;][]{UVOT1} on board the Neil Gehrels Swift Observatory. Its filter wheel can perform observations in the optical V,B and U bands and the ultraviolet UVW1, UVM2 and UVW2 bands. For the data reduction, we extracted the source counts with a radius of $3"-5"$ (depending on the apparent source extension and nearby sources), while the background counts are taken from a region around the source position avoiding visible contamination by the nearby star or other sources in the vicinity. Using the UVOTSOURCE tool provided by HEASOFT \footnote{For more details see \url{https://heasarc.gsfc.nasa.gov/docs/software/heasoft/}} we then computed the reddened magnitudes in for each observation in each filter.
Similarly, optical spectroscopic observations were collected from public archives and/or the literature for the six sources in the sample. In particular, for two sources we used the optical spectrum included in the 17th Data Release of the Sloan Digital Sky Survey \citep[SDSS-DR17][]{SDSS-DR17}, for one the Gran Telescopio Canarias (GTC) spectra publicly available on the ZBLLAC database \citep{ZBLLAC}, and for another one the Telescopio Nazionale Galileo (TNG) spectrum published in \cite{OlmoGarcia_2022}. The literature provided spectroscopic data for two additional sources \citep{Paliya_2021}. These observations are suited for estimating the physical properties of the central engine, constraining the parameter space exploration for the theoretical SED modeling, as explained in further detail in Section \ref{subsec: optical spectra}.

\subsection{X-ray data}\label{sec: x-ray}
In the context of this work, we first collected all available \textit{Swift}-X-Ray Telescope (XRT) \citep{Swift:2004,XRT:2005} observations for the sample. When available, we also retrieved data from other missions. In particular, two sources, 4FGL J1041.5+0607 and 4FGL J1040.5+0617, have an \textit{XMM-Newton} observation. In contrast, two sources do not have useful archived XRT data: 4FGL J2333.4-0133 and \J2115. For these cases, based on prior \textit{ROSAT} information \citep{ROSAT_ASS:1999,wgacat:1994}, we submitted two \textit{Swift} Targets of Opportunity (ToOs; ToO IDs: 23672 and 23673) to obtain a reliable flux measurement or a meaningful upper limit for SED modeling.

For the spectral modeling of the \textit{Swift}-XRT data, each individual pointing was processed using \texttt{FTOOLS} within the \texttt{HEASOFT} package (v6.33), following the same procedure described in \cite[][, hereafter \paperPHthree]{Zaballa:2025}. For the purpose of modeling the average SED of each source, when multiple observations were available, the filtered event files were summed using the \texttt{xselect} and \texttt{ximage} tools. Source and background spectra were then extracted in the 0.3--10.0 keV energy range. The ancillary response files and exposure maps were created with the \texttt{xrtmkarf} and \texttt{xrtexpomap} tasks, respectively.
For the \textit{XMM-Newton} data \citep{XMM:2001}, we followed the procedure described in \paperPHthree, using the XMM Science Analysis Software (SAS) to filter out time intervals affected by high background activity and to produce the corresponding spectral products for each EPIC instrument. When a source was detected by two or more instruments, the spectra were combined into a single spectrum, together with the corresponding response, ancillary, and background files, using the \texttt{epicspeccombine} task. Finally, all spectra were binned to a minimum of one count per spectral channel, enabling the use of C-statistics \citep{Cash:1979} in XSPEC. 
For all sources, the spectra are fitted in \texttt{XSPEC} considering an absorbed power law, fixing the value of Galactic absorption column density, using the Tuebingen-Boulder ISM absorption model \citep{Wilms:2000,XSPEC:1996}. 
\subsection{Gamma-ray data}\label{sec: gamma-ray}
High-energy (MeV-TeV) $\gamma$-ray observations are obtained from the \textit{Fermi} Large Area Telescope (\textit{LAT}) using data products published in the 4FGL-DR4 and the 4LAC-DR3 catalog \citep{ballet2024,4LAC_DR3:2022}, and the more recent \textit{Fermi}-LAT 16 year source list \citep[FL16Y;][]{FL16Y}. The catalogs provide both long-term variability information in the form of light curves and time-averaged SEDs, which are subsequently used for the broadband SED modeling. Additionally, we used the information regarding the $\gamma$-ray flux and corresponding spectral index, provided by the FL16Y, to estimate the total $\gamma$-ray luminosity, $\Lg$, following the approach described in \cite[][, hereafter \paperPHone]{Ghisellini_2009, Azzollini_2025} (see also Section \ref{subsec: optical spectra} for further details). For both spectra and light-curves we primarily used the FL16Y dataset. An exception is 4FGL J0127.3-0148, whose average significance falls below the FL16Y detection threshold and is therefore not included in the current source list. For this source, we instead use the information available in 4FGL-DR4 as it is a known gamma-ray emitter with high association probability $(>90\%)$ to its MWL counterpart.

\begin{figure}[h]
    \centering
    \includegraphics[width=\linewidth]{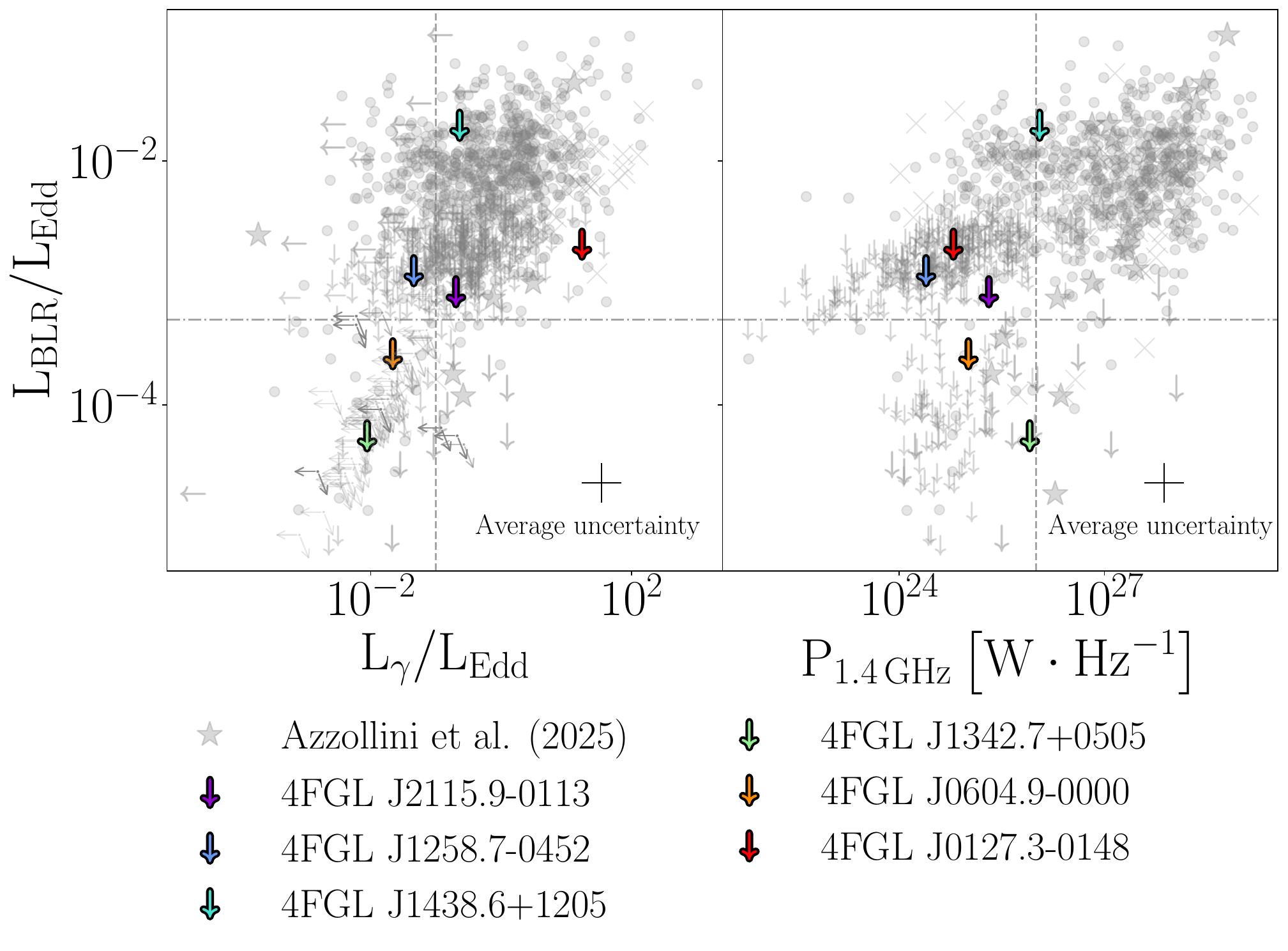}
    \caption{\label{fig: accretion vs. jet} Accretion regime and jet properties of the six sources in the target sample (see also Table \ref{tab: properties from optical}) compared to the blazar samples investigated in \paperPHone\ (gray). The horizontal and vertical gray dotted lines indicate the thresholds above which HERGs dominate over LERGs \citep[$\accretion\gtrsim5\times10^{-4}$, $\gratio\gtrsim0.1$, $\Pradio\gtrsim10^{26}\,\WHz$;][]{Sbarrato_2012, Ghisellini_2014, Heckman_Best:2014}. \textit{Left panel}: $\accretion$ vs. $\gratio$. \textit{Right panel}: $\accretion$ vs. $\Pradio$.}
\end{figure}

\section{Multifrequency properties of the sample} \label{sec: mwl}
\subsection{Physical properties derived from optical spectroscopy}
\label{subsec: optical spectra}

\begin{table*}[!h]
\centering
\caption{\label{tab: properties from optical} Primary sample and their physical properties derived in Section \ref{subsec: optical spectra}}
\resizebox{\textwidth}{!}{
\begin{tabular}{lllllcll}
\hline
4FGL Name & Redshift & $\mbh\left[\Msun\right]$ & $\accretion$ & $\gratio$ & $\Pradio\left[\WHz\right]$ & Class \\ 
\hline
J2115.9-0113            & $0.31$ & $<1.66\times10^{8}$ & $<8.47\times10^{-4}$ & $2.01\times10^{-1}$ & $2.03\times10^{25}$ & LERG$^1$          \\
J1258.7-0452            & $0.586^2$ & $<2.51\times10^{8}$ & $<1.26\times10^{-3}$ & $4.53\times10^{-2}$ & $2.45\times10^{24}$ & LERG$^2$        \\
J1438.6+1205            & $0.84772$ & $<1.79\times10^{6}$ & $<1.96\times10^{-2}$ & $2.29\times10^{-1}$   & $1.12\times10^{26}$ & HERG$^3$                  \\
J1342.7+0505            & $0.1366$ & $<3.63\times10^{8}$ & $<5.60\times10^{-5}$ & $8.71\times10^{-3}$ & $8.04\times10^{25}$ & LERG$^4$          \\
J0604.9-0000            & $0.30$ & $<7.61\times10^{8}$ & $<2.63\times10^{-4}$ & $2.16\times10^{-2}$ & $1.02\times10^{25}$ & LERG$^5$     \\
J0127.3-0148            & $0.33676$ & $<1.21\times10^{6}$ & $<2.08\times10^{-3}$ & $1.73\times10^{1}$   & $6.15\times10^{24}$ & LERG$^6$                 \\
\hline
\end{tabular}}
\textbf{Notes.}
\tablefoottext{1}{Redshift and physical properties estimated based on the optical spectrum from \cite{Paliya_2021}.}
\tablefoottext{2}{Redshift from \citet{Arsioli:2015}, physical properties estimated based on the optical spectrum from \cite{Padovani_2022}.}
\tablefoottext{3}{Redshift and physical properties estimated based on the optical spectrum from SDSS-DR17.}
\tablefoottext{4}{Redshift and physical properties estimated based on the optical spectrum from \cite{Paliya_2021}.}
\tablefoottext{5}{Redshift and physical properties estimated based on the optical spectrum from TNG.}
\tablefoottext{6}{Redshift and physical properties estimated based on the optical spectrum from SDSS-DR17.}
\end{table*}

Optical spectroscopic observations for the neutrino-emitter candidates in the target sample were collected from public archives and/or the literature as described in Section \ref{sec: observational data}. We used this dataset to infer the physical properties of the central engines of the blazars in the sample using the detected optical emission lines or placing limits in case of featureless spectra, following the procedure explained in \paperPHone. We were particularly interested in the luminosity of the broad-line region (BLR), of the accretion disk, and the black hole mass from which one can infer the Eddington luminosity $L_\mathrm{Edd} = 1.26\times 10^{38}\,(M_\mathrm{BH}/M_\odot)\,\mathrm{erg/s}$, as these properties probe the accretion regime of these sources. These are also key to discriminating blazar classes within a more physically driven classification scheme compared to the traditional, purely observational, taxonomy that distinguish them between BL Lacertae (BL Lacs) objects and flat-spectrum radio quasars (FSRQs) based on the presence and intensity of emission lines in the optical spectrum. 

A more physical scheme based on the modes of accretion and the intrinsic power of the relativistic jets has already been proposed as more effective in investigating the intrinsic nature of blazars in several works in the literature \citealt{Ghisellini_blue_FSRQ:2012, Sbarrato_2012, Heckman_Best:2014, Padovani_2022}, \paperPHone, \citealt{Azzollini_2025_CL}. Within this framework, the accretion regime properties are traced by the ratio of the luminosity of the BLR to the Eddington luminosity, $\accretion$, while the radio power at $1.4\,{\rm GHz}$, $\Pradio$, is a proxy for the intrinsic power of the jet. Furthermore, the $\gamma$-ray luminosity in Eddington units, $\gratio$, is also a good tracer of both the accretion efficiency (in the case of intense radiation fields) and the jet power (in all cases). Based on these properties, this more physically driven classification scheme distinguishes blazars into high-excitation radio galaxies (HERGs), which are characterized by the presence of intense radiation fields external to the jet, radiatively efficient accretion, and powerful jets, and low-excitation radio galaxies (LERGs), which typically show weak or no radiation fields, radiatively inefficient accretion, and reduced jet powers. The dividing lines between the two regimes lie at $\accretion\sim5\times10^{-4}$, $\gratio\sim0.1$, and $\Pradio\sim10^{26}\,\WHz$, with HERGs being characterized by all three properties above and LERGs dominating below the thresholds, respectively.
As previously done in \cite{Azzollini_2025_CL}, \paperPHone, we adopt this more physically driven HERG/LERG classification scheme to investigate the intrinsic properties of the six candidate neutrino-emitter blazars in the target sample. The corresponding estimated quantities are reported in Table \ref{tab: properties from optical}. The first column reports the 4FGL name of each source, while the second, third, fourth, fifth, and sixth list the redshift, the estimated black hole mass, accretion regime, jet power traced via the $\gamma$-ray luminosity, and radio power, respectively. The seventh column shows the resulting blazar class, while the corresponding reference for the optical spectrum used in the analysis is reported in the Table notes.
As reported in Table \ref{tab: properties from optical} and shown in Fig. \ref{fig: accretion vs. jet}, the six blazars in the target sample are characterized by featureless optical spectra in terms of broad emission lines (i.e., we could only place limits on the quantities of our interest), and predominantly show physical properties that are typical of LERG-like object ($\sim83\%$, i.e. five out of six sources).
Furthermore, the quantities reported in Table \ref{tab: properties from optical} were used as starting values to constrain the parameter space exploration for theoretical SED modeling significantly, as explained in further detail in the following Section.

\subsection{Building an SED}
Starting from the collected light curves (displayed for each source in Appendix \ref{Appendix: Lightcurves}) and spectral measurements (Section \ref{sec: observational data}), we construct time-averaged SEDs for each source in Table \ref{tab: properties from optical}. To reduce the impact or short-term variability and to avoid over-weighting densely sampled frequency ranges in the SED, the data are binned in logarithmic frequency with a bin width of $\Delta\log_{10}(\nu)=0.1$, where applicable. Within the bins the fluxes are averaged, and the flux error is propagated accordingly (Gauss). The multifrequency data are therefore combined as follows: radio measurements at low frequencies ($<10^{11}\,\mathrm{Hz}$) are treated as upper limits, as they are likely dominated by extended jet emission, not taken into account in the model. Infrared to ultraviolet (UV) data are derived from the procedure described above, while X-ray and gamma-ray spectra are taken as described in Sections \ref{sec: x-ray} and \ref{sec: gamma-ray} respectively. The resulting SEDs represent integrated source states over the full observation time available since $\sim2005$ and serve as the basis for the subsequent modeling.

\subsection{Multi-messenger modeling}
\label{subsec:model}
We consider a spherical emission region, located at a distance $R_\mathrm{diss}$ from the supermassive black hole, in a jet forming an angle $\theta_\mathrm{j}$ with the observer's line of sight. Particles in the emission region have a bulk Lorentz factor  $\Gamma_\mathrm{j}$, thus defining the Doppler factor as $\delta_\mathrm{j} = \frac{1}{\Gamma_\mathrm{j}(1 - \beta_\mathrm{j}\cos{\theta_\mathrm{j}})}$. For blazars we use $\theta_\mathrm{j}=1/\Gamma_\mathrm{j}$ and $\delta_\mathrm{j}=\Gamma_\mathrm{j}$.

The size of the emission region can be additionally constrained by the observed variability. Assuming that the light-crossing time of the photons in the emission region does not exceed the shortest variability time scale $t_{\mathrm{var}}$ observed in the light curves (see Appendix \ref{Appendix: Lightcurves} for details), this leads to a comoving blob radius
\begin{align}
    R'_{\mathrm{blob}}\lesssim\frac{c\ \delta_\mathrm{j}\ t_{\mathrm{var}}}{(1+z)}
\end{align}
In the case of the sample presented in this work, with redshifts ranging from $z=0.1366$ to $z=0.84772$ and for Doppler factors explored in a range between $\delta_\mathrm{j}=10$ and $\delta_\mathrm{j}=40$, we expect the emission region sizes to be $R'_{\mathrm{blob}}\lesssim10^{18}(t_\mathrm{var}/1\,\mathrm{yr})\,\mathrm{cm}$.

Accelerated particles (namely electrons and protons) are injected at a constant rate in the emission region, following a power law with an exponential cutoff. 
\begin{equation}
    Q'_i (E'_i) = Q_{i,0} E'^{-p_i}_i \exp{\left[-\left(\frac{E'_i}{E'_{i,max}}\right)^2\right]} \, \, \text{for $E'_i > E'_{min}$}
\end{equation}
such that $\int_{E_\mathrm{min}}^{\infty}Q' (E')E'dE' = L_{i,\mathrm{inj}}'$.

We use the software package\footnote{v1.2.0} AM$^3$ \citep{klinger_am3_2024} to compute the lepto-hadronic interactions in the emission region. The processes taken into account are synchrotron radiation, inverse-Compton scattering, Bethe-Heitler pair production, photo-pion production, and photon-photon annihilation. Particle escape is taken into account, with an escape time $t'_\mathrm{esc} = R'_\mathrm{jet}/c$, with $c$ the velocity of light. Details about the interactions, the computational methods and the software can be found in \cite{klinger_am3_2024}.

We evolve the simulation for each source until it reaches a quasi-steady state.
The interactions and densities are calculated in the blob's comoving frame, and we transform the resulting spectral flux (in units of energy per unit of surface per unit of time per unit of frequency) into the observer's frame as $F_\nu = (1+z) \, \delta_\mathrm{j}^3 \, L'_\nu / (4 \pi \, d_L^2)$, where $d_L$ is the luminosity distance, and the frequencies are related through $\nu = \delta_\mathrm{j} \nu' / (1+z)$.

Given the spectroscopic information on the Eddington ratio of the sources (see Section \ref{subsec: optical spectra}), we take into account photons from the accretion disk and the broad-line region, and we also consider the presence of a dusty torus in the vicinity of the supermassive black hole. The transformation of these external photon fields into the jet frame in order to carry out the radiative calculations follows \cite{ghisellini_madau_1996} and \cite{Ghisellini_2009}. Depending on the position of the emission region relative to the photons fields, parametrized as $r_\mathrm{diss} = R_\mathrm{diss}/R_\mathrm{BLR}$, the photons will be seen as boosted or de-boosted in the jet frame. 
Empirical relations are used to derive the characteristics of the external field, in order to reduce the number of free parameters in the models. Specifically, we use:
\begin{align*}
L_\mathrm{disk} &= 0.1 \, \dot M c^2,\\
L_\mathrm{BLR} &= 0.1 \, L_\mathrm{disk}\,\, \citep[\text{see e.g,}][]{ghisellini_madau_1996},\\
L_\mathrm{torus} &= f_\mathrm{torus}\,L_\mathrm{disk}, \\
\end{align*}
where $\dot{M}$ is the mass accretion rate of the disk, $L_\mathrm{disk}$, $L_\mathrm{BLR}$ and $L_\mathrm{torus}$ are the total disk, BLR and torus luminosities respectively and $f_\mathrm{torus}$ is the covering factor of the torus, that we fix to $f_\mathrm{torus} = 0.5$ \citep{calderone_widefield_2012, gu_evolution_2013}.
Furthermore, from e.g \cite{scaling_relation_sikora},
\begin{eqnarray}
R_\mathrm{BLR} &=& 10^{17} \, \left(\frac{L_\mathrm{disk}}{10^{45} \, \mathrm{erg/s}}\right)^{1/2} \, \mathrm{cm},\\
R_\mathrm{torus} &=& 1.2\times10^{18} \, \left(\frac{L_\mathrm{disk}}{10^{45} \, \mathrm{erg/s}}\right)^{1/2} \, \left(\frac{T_\mathrm{torus}}{10^3 \, \mathrm{K}}\right)^{-2}  \, \mathrm{cm}.
\label{eq:torus}
\end{eqnarray}
The torus temperature is in principle a free parameter, but it is limited by the dust sublimation temperature $T_\mathrm{torus} \lesssim 1300K$, and \cite{Cleary2007} showed that $T_\mathrm{torus} \sim 300 \, \mathrm{K}$. In order to limit the number of free parameters, we fix the torus temperature to $T_\mathrm{torus} = 300\,\mathrm{K}$.

The accretion disk (AD) is modeled as a multi-temperature blackbody, where we assume that the disk is perpendicular to the jet axis \citep{frank_accretion_2002}, and extending between the inner radius $R_\mathrm{in} = 6 \, G M_\mathrm{BH}/c^2 = 3 \, R_\mathrm{S}$, with $G$ the gravitational constant, where $R_\mathrm{g}$ is the Schwarzschild radius of the black hole and the outer radius $R_\mathrm{out} = 500 \,R_\mathrm{in}$: 
 \begin{equation*}
     F_\nu^\mathrm{AD}(\nu) =   \frac{4\pi h \nu^3}{c^2 d_L^2} \int_{R_\mathrm{in}}^{R_\mathrm{out}} \frac{R dR}{\exp{[h\nu/k_B T(R)] - 1}},
 \end{equation*}
where the temperature $T(R)$ is given by the standard relation for thin disks \citep[e.g.][]{frank_accretion_2002}
\begin{equation*}
T(R) = \left[\frac{3 G \dot{M} M_\mathrm{BH}}{8 \pi R^3 \sigma}\left(1 - \sqrt{\frac{R_\mathrm{in}}{R}}\right)\right]^{1/4},
\end{equation*}
$h$ is the Plank constant, $k_B$ is the Boltzmann constant and $\sigma$ is the Stefan-Boltzmann constant. 

The spectrum of the torus is described as a single-temperature blackbody, normalized to the luminosity of the dusty torus (DT) 
\begin{equation*}
    \nu\, F_\nu^\mathrm{DT}(\nu) = \frac{15 L_\mathrm{torus}}{4\pi^5 d_L^2} \left(\frac{h\nu}{k_B T_\mathrm{torus}}\right)^4 \left[\exp{\left(\frac{h\nu}{k_B T_\mathrm{torus}}\right) - 1}\right]^{-1}
\end{equation*}

In the case of \J2115, using an accretion rate $\dot{M} = 3.13\times 10^{-2} M_\odot/\mathrm{yr}$ corresponding to the upper limit of the Eddington ratio in Table \ref{tab: properties from optical}, would slightly overestimate the contribution of the disk to the total emission, as compared to the disk. For this reason, we  fix the accretion rate to $\dot{M} = 1.00\times 10^{-2} M_\odot/\mathrm{yr}$. With this value, the disk contribution is at most $\lesssim 10\%$ of the jet contribution. The rest of the sources' parameters related to the accretion, i.e., the black hole mass and the accretion rate, are fixed to the values from the upper limits in Table \ref{tab: properties from optical}.

Finally, some sources in the sample, for example GB6 J0604+0000, show a significant host galaxy contribution in their SED. While we do not aim to model the galactic emission of these sources in detail, we include a template for the host, taken from \cite{host_template}, that scales with the redshift and the black-hole mass \citep[we use the relation between the black-hole mass and the galaxy bulge mass from][]{host_scale} of each source, and remove the data points concerned by dominant host emission from the fitting procedure.

\section{Results}\label{sec: result}

\begin{figure*}
    \centering
    \includegraphics[width=\textwidth]{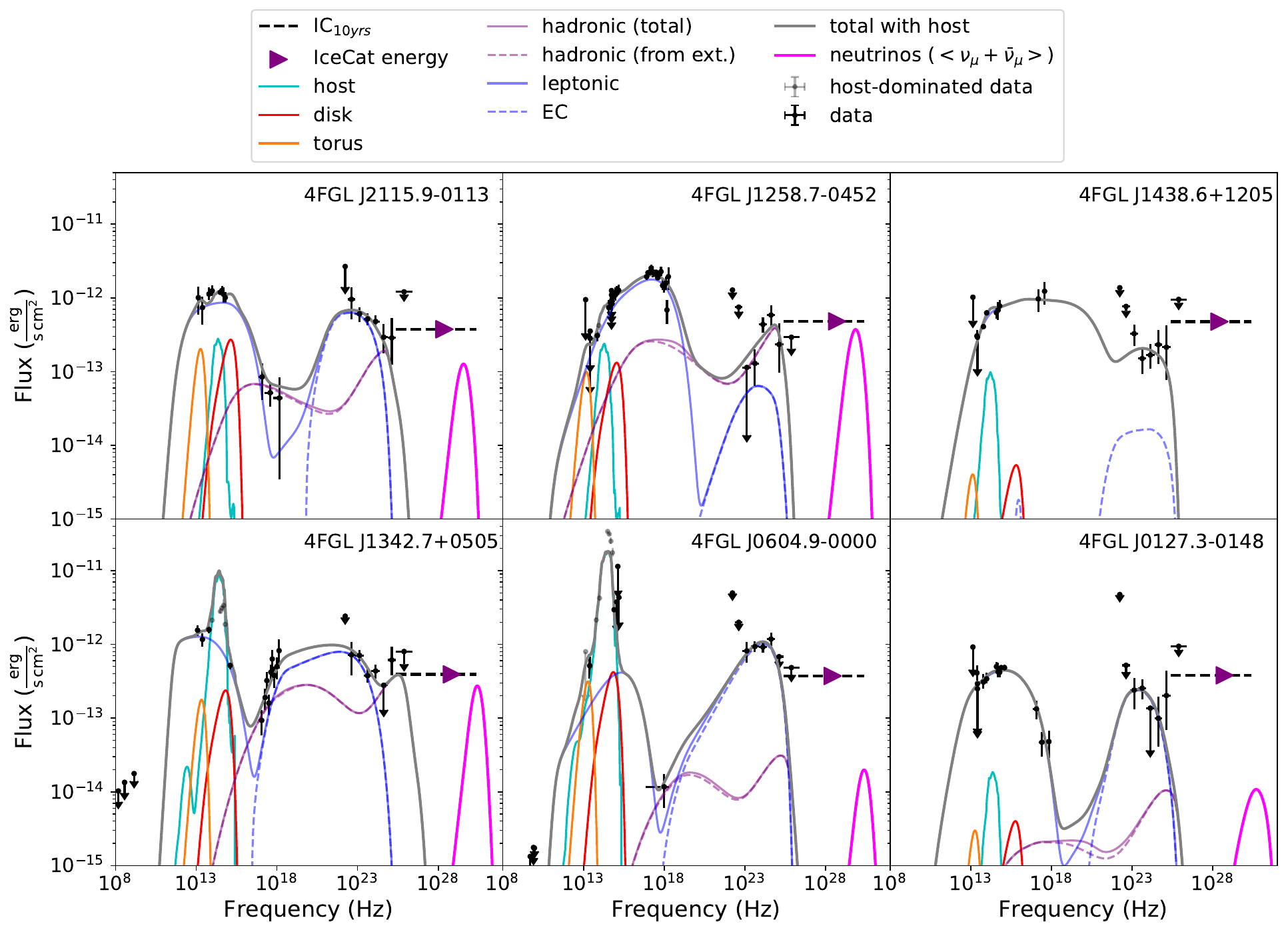}
    \caption{Lepto-hadronic best-fit solutions. The total emission is given by the gray line. We show the leptonic emission in blue, while the purple line shows the proton-initiated cascade emission (i.e including proton synchrotron emission and the synchrotron and inverse-Compton emissions of secondary particles from photon annihilation, Bethe-Heitler pair production and photo-pion pair production). The external-Compton (EC) component is highlighted with the dashed blue line, and the hadronic contribution originating from the interactions with the external photon fields is shown as a dashed purple line. The disk, torus, and host galaxy emissions are shown in red, orange, and cyan respectively. The muon and anti-muon neutrino flux is given by the pink curve, and compared to the 10-year IceCube sensitivity for each source (at its own declination). The estimated neutrino energy from the IceCAT catalog for each event is represented by a right-ward triangle (see text for a discussion on the neutrino peak energy).}
    \label{fig:lepto-hadronic_solutions}
\end{figure*}

Applying the model described in Section \ref{subsec:model}, we explore a set of parameter ranges, to find lepto-hadronic solutions that fit the time-averaged data, while when possible producing a non-negligible neutrino emission. To this aim, we make use of the genetic algorithm procedure \citep{genetic_algo} applied in \cite{rodrigues_2019,rodrigues_pks,Rodrigues:2024}. We use the algorithm to minimize the logarithmic $\chi^2$ cost function, then we check an additional condition on the resulting neutrino flux, namely we look for solutions that produce a steady-state neutrino flux larger than $10^{-14}\,\mathrm{erg\,cm^{-2}\,s^{-1}}$, and if this condition is not achieved we weight down the $\chi^2$ value (i.e. we set the $\chi^2$ value to $10^3$ in this case). Only in the case of one source is this condition not met while the SED fit is good, and we had to relax this condition on the neutrino output. The solutions are presented in Figure \ref{fig:lepto-hadronic_solutions} and the corresponding parameters can be found in Table \ref{tab:lepto-hadronic_solutions}. 

Individually, \J2115, \J1342, \J0604, and \J0127 exhibit typical lepto-hadronic SEDs, with a dominant leptonic component at low and high energies, and a subdominant contribution from the hadronic cascade, contributing mostly in the X-ray band. 

\begin{figure*}
    \centering
    \includegraphics[width=\textwidth]{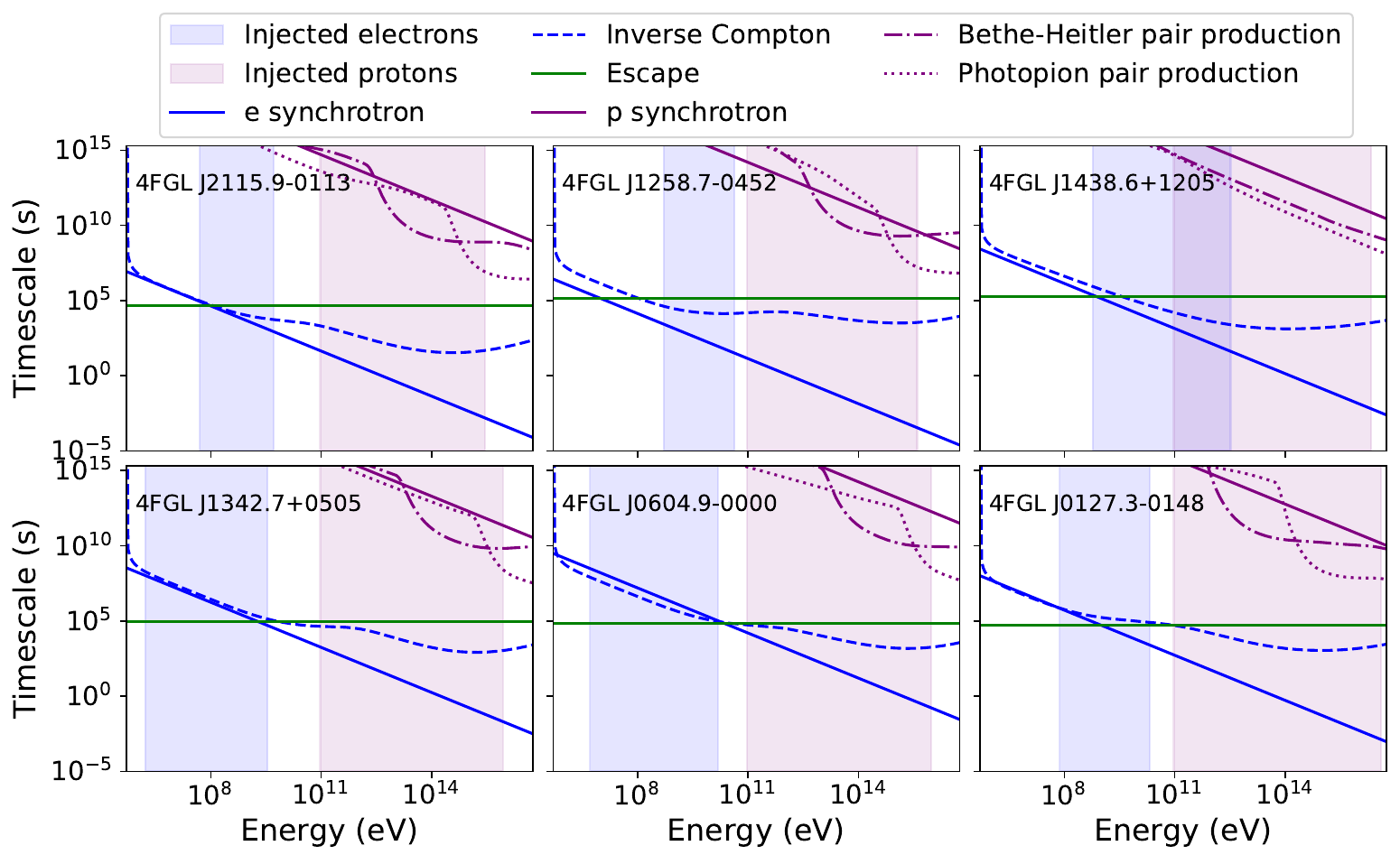}
    \caption{Loss timescales for the different processes concerning the primary particles (electron loss processes in blue and proton loss processes in purple). The escape timescale is represented by the green line, and is energy independent (see Section \ref{subsec:model}). The synchrotron, the inverse-Compton, the Bethe-Heitler pair production and the photo-pion pair production are shown with the solid, dashed, dash-dotted and dotted lines, respectively. The shaded areas indicate the range of energy at which the primary electrons and protons are continuously injected.}
    \label{fig:timescales}
\end{figure*}

The high-energy part of the SED of \J1342 challenges the model. The very hard X-ray spectrum is difficult to reconcile with the flatter gamma-ray spectrum through only the inverse-Compton component that drops sharply after $\sim 10^{23}\, \mathrm{Hz}$. The additional contribution of the hadronic cascade helps to ease the tension. In Figure \ref{fig:timescales}, we present the loss timescales $(t_\mathrm{loss} = \left[E^{-1}\left(\frac{dE}{dt}\right)_\mathrm{loss}\right]^{-1})$ for the different processes involved in the modeling. We show the energy range at which primary particles are injected with the shaded areas. Using that, we see that for the SEDs of \J1342 and \J0604, while looking similar to the others', they differ by the dominant loss processes. For these two sources, primary electrons tend to escape faster than they cool by emitting synchrotron radiation, over most of the energies of the injected distribution (still enabling a quasi steady-state to set in). On the other hand, in the other sources, the resulting leptonic emission originates from the balance between synchrotron emission and escape, creating a quasi steady-state.

The solution for \J1438 is entirely dominated by leptonic emission, and the hadronic contributions are not visible in Figure \ref{fig:lepto-hadronic_solutions}, as they peak below the spectral upper limits of the plot. This extreme sub-dominance is due to the fact that we do not consider solutions that have a physical injected proton luminosity above the Eddington luminosity -- which for this source is low, given the black-hole mass of the order of $10^6 \, M_\odot$ --, but also to the lower impact of the external fields. Indeed, with $\Gamma_\mathbf{j} = 16.2$ and $R_\mathrm{diss} = 2.8\, R_\mathrm{BLR}$ the boosting of the photon fields into the blob's frame is reduced compared to \J0127 which shares similar properties but for a larger Lorentz factor and thus a larger neutrino flux (even though much lower than the IceCube sensitivity). These effects are worsened by the larger redshift of the source compared to the other sources in the sample, because of the larger luminosity distance while transforming to the observer's frame. The source \J1258 is best modeled with a dominant contribution at high energies from the hadronic cascade, which is able to explain the peaky feature in gamma rays. 

\begin{table*}
\caption{Parameters of the lepto-hadronic models.}
\begin{adjustbox}{width=\textwidth}
    \centering
    \begin{tabular}{c|c|c|c|c|c|c|c}
        \hline
         & unit & \J2115 & \J1258 & \J1438 & \J1342 & \J0604 & \J0127\\
        \hline
        $R'_{\rm blob}$ & cm & $1.31 \times 10^{15}$ & $4.15 \times 10^{15}$ & $5.66 \times 10^{15}$ & $2.81 \times 10^{15}$ & $2.04 \times 10^{15}$ & $1.66 \times 10^{15}$ \\
        $B'$ & G & $9.55$ & $1.70 \times 10^{1}$ & $1.70$ & $1.50$ & $5.00 \times 10^{-1}$ & $2.75$ \\
        $\Gamma_{\rm blob}$ & - & $1.41 \times 10^{1}$ & $2.48 \times 10^{1}$ & $1.62 \times 10^{1}$ & $2.52 \times 10^{1}$ & $2.55 \times 10^{1}$ & $2.62 \times 10^{1}$ \\
        $\gamma'_{\rm e,min}$ & - & $9.82 \times 10^{1}$ & $9.90 \times 10^{2}$ & $1.15 \times 10^{3}$ & $3.20$ & $1.00 \times 10^{1}$ & $1.48 \times 10^{2}$ \\
        $\gamma'_{\rm e,max}$ & - & $1.00 \times 10^{4}$ & $8.53 \times 10^{4}$ & $6.46 \times 10^{6}$ & $6.87 \times 10^{3}$ & $3.00 \times 10^{4}$ & $4.06 \times 10^{4}$ \\
        $p_{\rm e}$ & - & $2.00$ & $1.50$ & $2.00$ & $2.80$ & $2.20$ & $2.28$ \\
        $\gamma'_{\rm p,max}$ & - & $3.00 \times 10^{6}$ & $4.22 \times 10^{6}$ & $2.32 \times 10^{7}$ & $9.16 \times 10^{6}$ & $1.00 \times 10^{7}$ & $4.32 \times 10^{7}$ \\
        $L'_{\rm e}$ & erg/s & $4.01 \times 10^{40}$ & $1.90 \times 10^{40}$ & $3.00 \times 10^{41}$ & $8.00 \times 10^{40}$ & $3.20 \times 10^{40}$ & $3.35 \times 10^{39}$ \\
        $L'_{\rm p}$ & erg/s & $2.50 \times 10^{43}$ & $2.10 \times 10^{43}$ & $1.72 \times 10^{42}$ & $1.50 \times 10^{43}$ & $1.00 \times 10^{43}$ & $3.92 \times 10^{41}$ \\
        $R_{\rm diss}$ & $R_{\rm BLR}$ & $1.04$ & $1.30$ & $2.80$ & $1.80$ & $1.90$ & $1.48$ \\
        \hline
        $\dot{M}$ & $M_\odot/{\rm yr}$ & $3.13 \times 10^{-2}$ & $7.04 \times 10^{-2}$ & $7.00 \times 10^{-3}$ & $4.59 \times 10^{-3}$ & $4.50 \times 10^{-2}$ & $5.60 \times 10^{-4}$ \\  
        $L_{\rm disk}$ & erg/s & $1.77 \times 10^{44}$ & $3.99 \times 10^{44}$ & $3.96 \times 10^{43}$ & $2.60 \times 10^{43}$ & $2.55 \times 10^{44}$ & $3.17 \times 10^{42}$ \\  
        $M_{\rm BH}$ & $M_\odot$ & $1.66 \times 10^{8}$ & $2.51 \times 10^{8}$ & $1.79 \times 10^{6}$ & $3.63 \times 10^{8}$ & $7.61 \times 10^{8}$ & $1.21 \times 10^{6}$ \\  
        $R_{\rm diss}$ & cm & $4.38 \times 10^{16}$ & $8.21 \times 10^{16}$ & $5.57 \times 10^{16}$ & $2.90 \times 10^{16}$ & $9.59 \times 10^{16}$ & $8.33 \times 10^{15}$ \\  
        $R_{\rm torus}$ & cm & $5.61\times 10^{18}$ & $8.42\times 10^{18}$ & $2.65\times 10^{18}$ & $2.15\times 10^{18}$ & $6.73\times 10^{18}$ & $7.51\times 10^{17}$ \\  
        \hline
        $U'_{\rm e}/U'_{\rm B}$ &  & $1.71 \times 10^{-2}$ & $2.55 \times 10^{-4}$ & $2.16 \times 10^{-1}$ & $2.99 \times 10^{-1}$ & $2.05$ & $1.07 \times 10^{-2}$ \\
        $U'_{\rm p}/U'_{\rm B}$ &  & $1.07 \times 10^{1}$ & $2.82 \times 10^{-1}$ & $1.24$ & $5.60 \times 10^{1}$ & $6.41 \times 10^{2}$ & $1.26$ \\
        \hline
        $N_{<\nu_\mu + \bar{\nu}_\mu>}$ &  & $2.60 \times 10^{-2}$ & $4.90 \times 10^{-2}$ & $1.54 \times 10^{-5}$ & $1.69 \times 10^{-2}$ & $1.82 \times 10^{-3}$ & $1.32 \times 10^{-3}$ \\
        CI (10 yr) & - & $0-2$ & $0-3$ & $0-0$ & $0-2$ & $0-1$ & $0-1$ \\
        \hline
        $\chi^2_\mathrm{red} \, (\mathrm{d.o.f})$ & - &$ 1.7\,(7)$&$1.8\,(17)$&$ 3.1\,(3)$&$ 3.6\,(9)$&$ 7.4\,(1^*)$&$2.7\,(5)$\\
        \hline
    \end{tabular}
    
\end{adjustbox}
\textbf{Notes.} The proton spectral index is fixed to $p_\mathrm{p}= 1.5$ and the proton minimum Lorentz factor to $\gamma'_\mathrm{p,min} = 100$, since with this spectral index, the value does not impact the results much. ($^*$): The number of free parameters for this SED is larger than the number of data points when accounting for the host contribution ($\mathrm{d.o.f}< 0$), so in this case we are effectively over-fitting, strictly speaking, and use a number of degrees of freedom of $1$.
    \label{tab:lepto-hadronic_solutions}
\end{table*}
We find that compact emission regions, with a size of the order of a few $10^{15}\,\mathrm{cm}$, located within the BLR can explain the SEDs, with $R_\mathrm{diss} = (1.04 -2.80)\,R_\mathrm{BLR}$. This corresponds to distances much below the parsec scale, given the BLR size in these sources. This location leads to a strong external Compton component at high energies, while enhancing the hadronic cascade and supporting neutrino production. This effect is further intensified by the large Doppler factor inferred from the best-fit solutions. We find values between $14$ and $26$ for the Lorentz factor, which increases the boosting of the external field in the comoving frame of the jet, allowing them to be preferential targets for photo-pion production and neutrino production.

We find solutions that are close to equipartition between the energy density $U'$ of relativistic particles and magnetic field ($U'_\mathrm{particles}/U'_\mathrm{B} \sim 0.3-10$ for \J2115, \J1258, \J1438 and \J0127), or more particle dominated ($U'_\mathrm{particles}/U'_\mathrm{B} \sim 50 - 640$ for \J1342 and \J0604), with magnetic field strength between $5 \times 10^{-1} \, \mathrm{G}$ and $1.7\times 10^1 \, \mathrm{G}$, with electron spectral indices $p_\mathrm{e} = 2.0 - 2.8$, except for \J1258 where the electrons are injected with the same spectral index as the protons, namely $1.5$. 
The magnetic field strengths are typical from lepto-hadronic models \cite[see, e.g.][]{Mastichiadis_2013}, and consistent with estimates from radio observations \citep{Pushkarev_2012}. Assuming the average values at $1\,\mathrm{pc}$ from \cite{Pushkarev_2012}, $B'_\mathrm{1pc} \sim 0.4 - 0.9\, \mathrm{G}$ and a magnetic field strength evolution such that $B'(r) = B'_\mathrm{1pc}\left(\frac{r}{1\mathrm{pc}}\right)^{-1}$ \citep{Lobanov_1998}, we find values at the location of the emission regions ($R_\mathrm{diss} \sim 8\times 10^{15} - 10^{17}\, \mathrm{cm}$) of the order of $10-100\,\mathrm{G}$, the latter being extreme for lepto-hadronic models.

We also report the number of muon and anti-muon neutrinos $N_{<\nu_\mu + \bar{\nu}_\mu>}$ detectable by IceCube per year predicted by the models in the last row, calculated as: 
\begin{eqnarray*}
    N_{<\nu_\mu + \bar{\nu}_\mu>} = \Delta t \int_{E_\mathrm{min}}^{E_\mathrm{max}} dE \, A_\mathrm{eff}(E,\delta) \, \Phi_{<\nu_\mu + \bar{\nu}_\mu>}(E)
\end{eqnarray*}
where $\Phi_{<\nu_\mu + \bar{\nu}_\mu>}(E)$ is the sum of the muon neutrino and anti-neutrino flux in units of neutrinos per unit energy per unit of time per unit of surface, predicted by the models, $A_\mathrm{eff}(E,\delta)$ is the effective area of the detector\footnote{available at \url{https://IceCube.wisc.edu/data-releases/2021/01/all-sky-point-source-IceCube-data-years-2008-2018/}}, at the position of the source and $\delta$ is the declination of the source. 

The predicted neutrino emissions of the sources peak above PeV energies, which is always larger than the estimated energy reported in IceCAT. However, the reconstructed energy reported in the IceCAT catalog is calculated assuming a $E_\nu^{-2.19}$ astrophysical neutrino power law flux \citep{IceCAT-1:2023}. While assuming a harder spectral shape, as predicted by the models supported by a photo-pion origin of the neutrinos, the IceCube estimated energy may actually lie above the IceCAT value \citep{kuhlmann_capel_2025,rodrigues_neutrinos_review_2026}. The predicted neutrino rates vary from $10^{-5}$ to $5\times 10^{-2}$ neutrino per year. In this regard, the most promising source is \J1258. For each source, we also compute the 99\% confidence interval (CI in Table \ref{tab:lepto-hadronic_solutions}) of fluctuation of the neutrino rate, assuming it follows a Poisson statistics. The sources \J0604 and \J0127 have confidence intervals expected between $[0-1]$ assuming ten years of observation. For \J2115, \J1258, and \J1342 the confidence intervals vary between $[0-2]$, $[0-2]$ and $[0-3]$, respectively, and for the least promising source (\J1438), we do not predict any significant neutrino emission. As mentioned earlier, this results from the combination of restricting the proton injected luminosity and a lower bulk Lorentz factor.

\J1438, the only HERG in the sample, is the least efficient neutrino emitter, despite the relative strength of its accretion disk. In fact, the two least promising sources with respect to neutrino prediction have the smallest black hole masses, which drastically limits the amount of injected protons that can contribute to the neutrino emission if the jet power is assumed to stay below or at the level of the Eddington luminosity.

\section{Discussion}\label{sec: discussion}

\begin{figure*}
    \centering
    \includegraphics[width=0.8\linewidth]{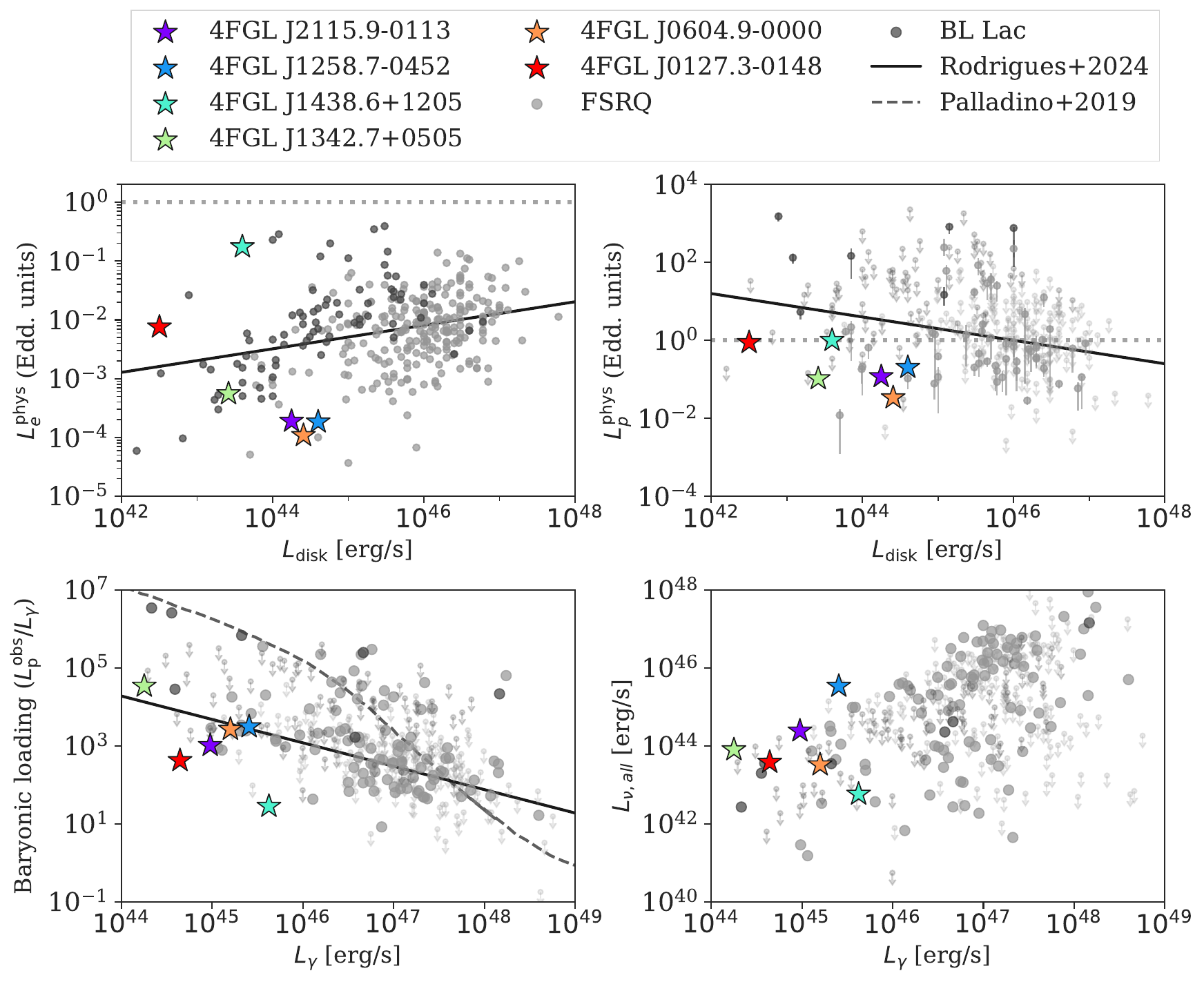}
    \caption{Comparison of the sample with the 324 blazars from \cite{Rodrigues:2024}. Top panels: physical electron (left) and proton (right) luminosities  normalized to the Eddington luminosity of each source, as a function of their disk luminosity. Bottom left: baryonic loading  as a function of the gamma-ray luminosity. Bottom right: all-flavor neutrino luminosity predicted by the models, as a function of the gamma-ray luminosity. The sample is shown by the colored star markers, while the sample modeled by \cite{Rodrigues:2024} is shown in the background as gray points and upper limits.}
    \label{fig:sample_comparison_rodrigues}
\end{figure*}

We compare the best-fit solutions of our sample to the 324 blazars modeled in \cite{Rodrigues:2024}. In Figure \ref{fig:sample_comparison_rodrigues} we show different characteristics of the sample. Namely, the ratio of proton and electron physical luminosities ($\Gamma_\mathrm{j}^2L'_\mathrm{e,p}/2$) to the Eddington luminosity as a function of the disk luminosity, the baryonic loading ($L^\mathrm{obs}_\mathrm{p}/L_\gamma$, where $L^\mathrm{obs}_\mathrm{p} = \Gamma_\mathrm{j}^4\,L'_\mathrm{p}$) and the all-flavour neutrino luminosity $L_{\nu,\mathrm{all}}$, as a function of the gamma-ray luminosity (integrated between $100\,\mathrm{MeV}$ and $100\,\mathrm{GeV}$).

Our sample contains five LERGs and one HERG. When compared to the blazar sample of \cite{refId0}, the properties of our models consistently align with their BL Lac objects in Figure \ref{fig:sample_comparison_rodrigues}. Even for the HERG source \J1438 in which the accretion efficiency traced by the ratios $L_\mathrm{BLR}/L_\mathrm{Edd}$ and $L_\gamma/L_\mathrm{Edd}$ (see Section \ref{subsec: optical spectra}) is larger than for the other sources in the sample, the modeled properties lie close to the BL Lacs in the sample from \cite{refId0}. This emphasizes on the limitation of the BL Lac/FSRQ classification. The spread in electron luminosity is explained by the intrinsic differences among the modeled SEDs. On the other hand, the proton luminosities vary between $3\times 10^{-2} \, L_\mathrm{Edd}$ and $L_\mathrm{Edd}$, with the upper bound imposed on the models in order to keep the jet powers sub-Eddington.
Given the baryonic loading of the sources ($2\times10^1-4\times10^4$), they are unlikely to be contributing significantly to the IceCube diffuse flux. This can be seen in the bottom left panel of Figure \ref{fig:sample_comparison_rodrigues}, where the baryonic loading of the sources are several orders of magnitude below the limiting baryonic loading determined by \cite{Palladino:2019} in order not to overshoot the IceCube stacking limit \citep{IceCube2017_2LAC}. 
The all-flavor neutrino luminosity spreads over three orders of magnitude, and does not show any specific trend with the gamma-ray luminosity, since for five out of six sources in the sample, gamma rays originate from inverse-Compton scattering (Figure \ref{fig:lepto-hadronic_solutions}). The neutrino production parameter defined as $Y_{\nu\gamma} = L_{\nu_\mu}/L_\gamma$ varies between $4\times 10^{-4}$ for \J1438 and $4\times10^{-1}$ for \J1258. This parameter is another way of looking at the lower right panel of Figure \ref{fig:lepto-hadronic_solutions}. We do not find a correlation between gamma-ray luminosity and neutrino luminosity, because gamma-rays are predominantly of leptonic origin (except for \J1258). Several studies have shown that $Y_{\nu\gamma} \sim 5 - 10\%$ for blazars  given that blazars are not found to dominate the contribution to the diffuse neutrino flux \citep[e.g][]{Aartsen_20172LAC, Wang_Loeb_nu_bkg:2016, Padovani_simplified:2015, capel_assessing_2022}.

Despite being dominantly LERGs, the sources nonetheless benefit from the presence of their external fields, as seen when comparing dashed and solid lines in Figure \ref{fig:lepto-hadronic_solutions}, due to location of the emission region. The contribution of the external fields is further enhanced by the large Lorentz factor predicted by the models. In the most efficient neutrino emitter of the sample, \J1258, the effect of including external fields can be seen when comparing the results to those of \cite{refId0}. In that work, the authors modeled the sources without the contribution from external fields, due to the BL Lac classification of the source. While they were able to find a good fit to the SED, we cannot compare directly the chi-square value obtained with their fitting procedure, as it is different from the one in this work. Despite this, the neutrino output is about a factor 5 lower than what we find while being conservative in the energy budget of the jet, and having a reduced (linear) chi-square value below $2$ in this study. With the inclusion of the external fields upper limits from Table \ref{tab: properties from optical}, the dominant contribution is switched to the hadronic cascade in our case, compared to \cite{refId0}, where it only played a sub-dominant role. In fact, the hadronic fraction in our case explains $90\%$ of the total peak flux in gamma-rays.

The models assume that particles are already accelerated when they are injected into the emission region. The power-law slope and magnetic field strength values are consistent with diffusive shock acceleration of electrons, with soft indices $p_\mathrm{e} = 2.00 - 2.80$ and regions that are particle dominated (in the case of \J0604 and \J1342) or near equipartition (for \J2115, \J1438, and \J0127) $U'_\mathrm{particles}/U'_\mathrm{B} \sim 1.30 - 6.41 \times 10^2$ \citep[see e.g][for prediction of the particles spectrum undergoing shock acceleration]{DSA_low_mag,DSA_unmagnetized}. One exception is \J1258, for which the energy density ratio $U'_\mathrm{particles}/U'_\mathrm{B} \sim 3\times 10^{-1}$ and the electron spectral index is hard, which hints in this case toward an acceleration through magnetic reconnection \citep[see e.g][]{magnetic_reco,hardo_slope_reco}.With this assumption, different mechanisms could accelerate the protons and the electrons separately. However, if instead we assume that the same process accelerates both populations of primary particles \citep[co-acceleration assumption, see e.g][for an application to shock acceleration]{co_acceleration}, the spectral index $p_\mathrm{p}$ should be similar the spectral index of the electrons $p_\mathrm{e}$ and in order to retrieve the same predicted neutrino flux and hadronic cascade while keeping the rest of the parameters fixed, this would require larger (up to a factor 100 larger) injected proton luminosities, in tension with a sub-Eddington jet power for the average SEDs. If we assume a conservative proton spectral index $p_\mathrm{p}=2$, consistent with DSA but not necessarily equal to $p_\mathrm{e}$, for the all five sources which have softer electron spectral index, the tension decreases, with jet powers still below the Eddington luminosity except for the two low-BH-mass sources \J1438 and \J0127. The only source not concerned by this limitation is \J1258, for which the electrons are already found to be injected with the same spectral shape as the protons.

We note that even if we do not constrain the relativistic particles to be in equipartition with the magnetic field, we find four solutions out of six within an order of magnitude from equipartition. As discussed in \cite{2013ApJ...768...54B}, this is consistent with the current understanding of jet launching via the Blandford-Znajek mechanism \citep{Blandford_Znajek_1977} (where the jet is launched as magnetically-dominated and converts the magnetic power to kinetic power), although it can be challenging to reconcile with SED modeling \citep[see, e.g.][]{2013ApJ...768...54B, Petropoulou_2017, Zech_2017, Sol_2022}. Here the solutions benefit from large magnetic field strengths while at the same time the high-energy band is accounted for by the interactions with the external photon fields, and the proton energy density is limited by the constraint on the jet power, leading to solutions that do not depart significantly from equipartition.
For the most extreme solution that display a particle-dominated composition, i.e \J0604, the tension could be softened if the proton luminosity is decreased, since the hadronic contribution is not dominant in the SED model.

\section{Conclusion}
In conclusion, we have applied a lepto-hadronic model to a sample of six blazars composed of five LERGs and one HERG, taking into account the influence of external photon fields based on spectroscopical limits on the redshift, black-hole mass and accretion disk luminosity. From this model we derived a best-fit solution, focused on maximizing the neutrino output, for each source and calculate the predicted rate of neutrinos detectable by IceCube. 

On average the sources in the sample have low disk luminosities and low gamma-ray luminosities, but the location of the emission region and the Doppler boosting of the jet supports the production of neutrinos via interactions with the external photon fields, with the exception of \J1438, for which we do not predict any significant neutrino emission, due to a moderate Lorentz factor, a larger redshift and the limitation of the injected proton luminosity to retrieve jet powers smaller than the Eddington luminosity of the source. The relativistic particles are close to equipartition with the magnetic field for four of the six sources, and we ensure the total jet power remains below the Eddington luminosity.

The most efficient neutrino emitter of the sample is \J1258, for which we predict an observable neutrino rate of $5\times 10^{-2}/\mathrm{yr}$. While these blazars are not expected to be significantly contributing to the diffuse flux observed by the IceCube neutrino observatory given their baryonic loading, in five out of six cases they are compatible with the observation of one IceCube high-energy event when accounting for Poisson statistics.

Even if the gamma-rays and the neutrinos are traced each by a different primary particle population in the sample (again, except in the most promising case of \J1258), the external photon fields are a crucial ingredient of the modeling of these sources. In this sense, this study demonstrates that the limits placed on the physical properties of the sources, such as the disk luminosity or the black-hole mass, need to be taken into account in the context of neutrino emission predictions.

\section{Data availability}
The SED data for the six blazars modeled in this work are publicly available on Zenodo at \url{https://doi.org/10.5281/zenodo.22799753}.
\begin{acknowledgements}
The authors thank Andrew Taylor and Xavier Rodrigues for useful discussions and feedback on the manuscript.

This work was supported by the European Research Council, ERC Starting grant \emph{MessMapp}, S.B. Principal Investigator, under contract no. 949555. 

S.G. is grateful for the support of the Koshland Family Foundation.

This research has made use of data and/or software provided by the High Energy Astrophysics Science Archive Research Center (HEASARC), which is a service of the Astrophysics Science Division at NASA/GSFC.

This research furthermore has made use of the NASA/IPAC Extragalactic Database, which is funded by the National Aeronautics and Space Administration and operated by the California Institute of Technology.

Part of this work is based on archival data, software or online services provided by the Space Science Data Center - ASI.

This research has made use of public data provided by the $Fermi$ Science Support Center (FSSC) maintained by the High Energy Astrophysics Science Archive Research Center (HEASARC) at NASA Goddard Space Flight Center."

We acknowledge the use of public data from the Neil Gehrels Swift Observatory data archive.

This work presents results from the European Space Agency (ESA) space mission Gaia. Gaia data are being processed by the Gaia Data Processing and Analysis Consortium (DPAC). Funding for the DPAC is provided by national institutions, in particular the institutions participating in the Gaia MultiLateral Agreement (MLA). The Gaia mission website is \url{https://www.cosmos.esa.int/gaia}. The Gaia archive website is \url{https://archives.esac.esa.int/gaia}.

This work is based on observations obtained with the Samuel Oschin Telescope 48-inch and the 60-inch Telescope at the Palomar Observatory as part of the Zwicky Transient Facility project. ZTF is supported by the National Science Foundation under Grant No. AST-2034437 and a collaboration including Caltech, IPAC, the Weizmann Institute for Science, the Oskar Klein Center at Stockholm University, the University of Maryland, Deutsches Elektronen-Synchrotron and Humboldt University, the TANGO Consortium of Taiwan, the University of Wisconsin at Milwaukee, Trinity College Dublin, Lawrence Livermore National Laboratories, and IN2P3, France. Operations are conducted by COO, IPAC, and UW.

This work has made use of data from the Asteroid Terrestrial-impact Last Alert System (ATLAS) project. The Asteroid Terrestrial-impact Last Alert System (ATLAS) project is primarily funded to search for near earth asteroids through NASA grants NN12AR55G, 80NSSC18K0284, and 80NSSC18K1575; byproducts of the NEO search include images and catalogs from the survey area. This work was partially funded by Kepler/K2 grant J1944/80NSSC19K0112 and HST GO-15889, and STFC grants ST/T000198/1 and ST/S006109/1. The ATLAS science products have been made possible through the contributions of the University of Hawaii Institute for Astronomy, the Queen’s University Belfast, the Space Telescope Science Institute, the South African Astronomical Observatory, and The Millennium Institute of Astrophysics (MAS), Chile.

We acknowledge the use of public data from the Catalina Real-Time Transient Survey.

The Pan-STARRS1 Surveys (PS1) and the PS1 public science archive have been made possible through contributions by the Institute for Astronomy, the University of Hawaii, the Pan-STARRS Project Office, the Max-Planck Society and its participating institutes, the Max Planck Institute for Astronomy, Heidelberg and the Max Planck Institute for Extraterrestrial Physics, Garching, The Johns Hopkins University, Durham University, the University of Edinburgh, the Queen's University Belfast, the Harvard-Smithsonian Center for Astrophysics, the Las Cumbres Observatory Global Telescope Network Incorporated, the National Central University of Taiwan, the Space Telescope Science Institute, the National Aeronautics and Space Administration under Grant No. NNX08AR22G issued through the Planetary Science Division of the NASA Science Mission Directorate, the National Science Foundation Grant No. AST-1238877, the University of Maryland, Eotvos Lorand University (ELTE), the Los Alamos National Laboratory, and the Gordon and Betty Moore Foundation.

This publication makes use of data products from the Wide-field Infrared Survey Explorer, which is a joint project of the University of California, Los Angeles, and the Jet Propulsion Laboratory/California Institute of Technology, funded by the National Aeronautics and Space Administration.\\
\end{acknowledgements}

\bibliographystyle{aa} 
\bibliography{aa62198-26_arxiv}

\begin{appendix}
\counterwithin{figure}{section}
\onecolumn
\section{Multifrequency light curves}
\label{Appendix: Lightcurves}
\begin{figure*}[h]
    \centering
    \includegraphics[width=1\linewidth]{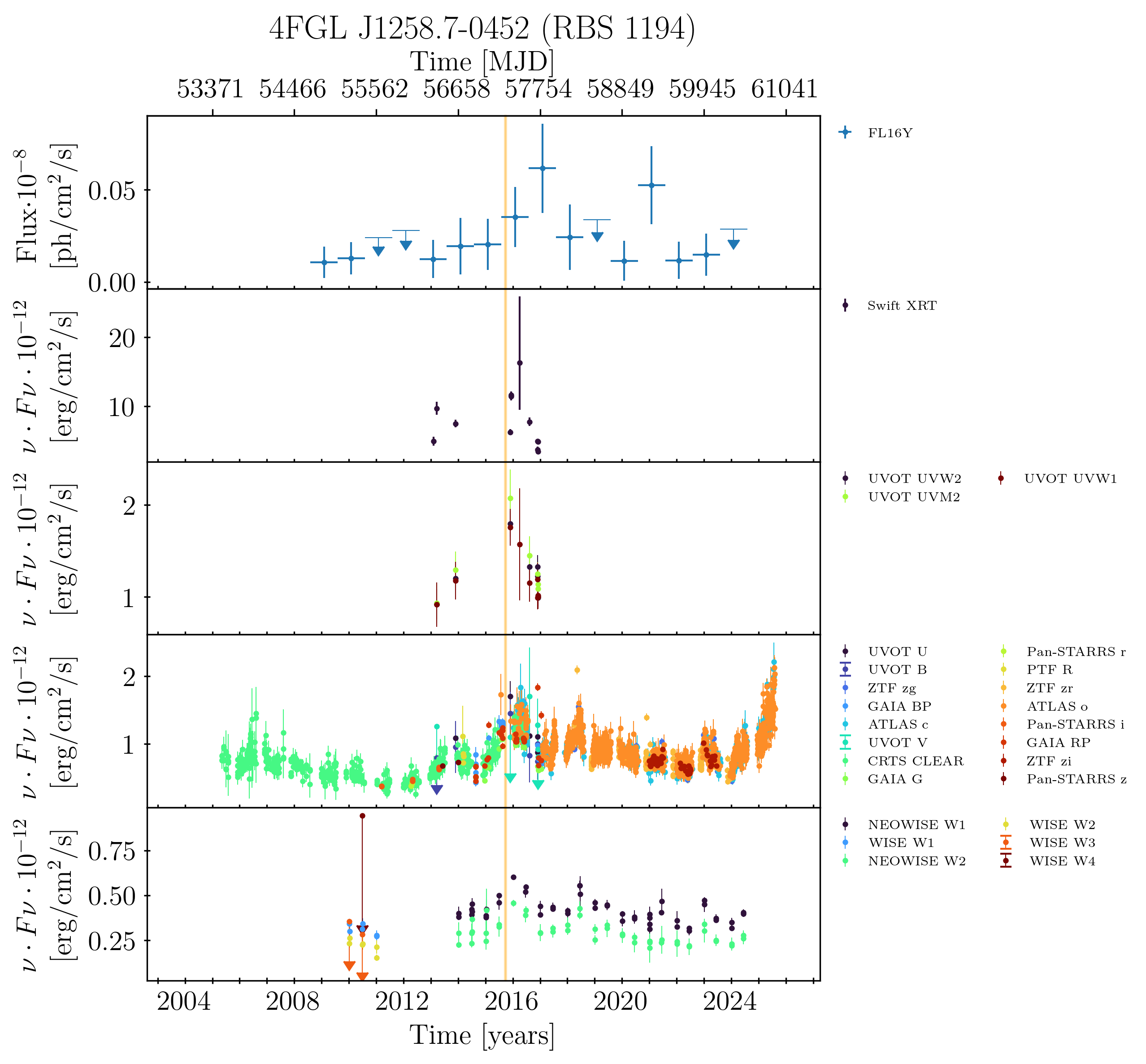}
    \caption{Multifrequency light curves of \J1258. The panels display from top to bottom: \textit{Fermi}-LAT gamma-ray light curves integrated over a 12-months time bin taken from the FL16Y source list; X-ray data from \textit{Swift-}XRT; UV data from \textit{Swift-}UVOT; optical data from different facilities; IR data from WISE/NEOWISE. The orange stripe highlights the arrival time of the neutrino candidate associated with this source (IC150926A).}
    \label{fig:LC_J1258m0447}
\end{figure*}

\begin{figure*}[h]
    \centering
    \includegraphics[width=1\linewidth]{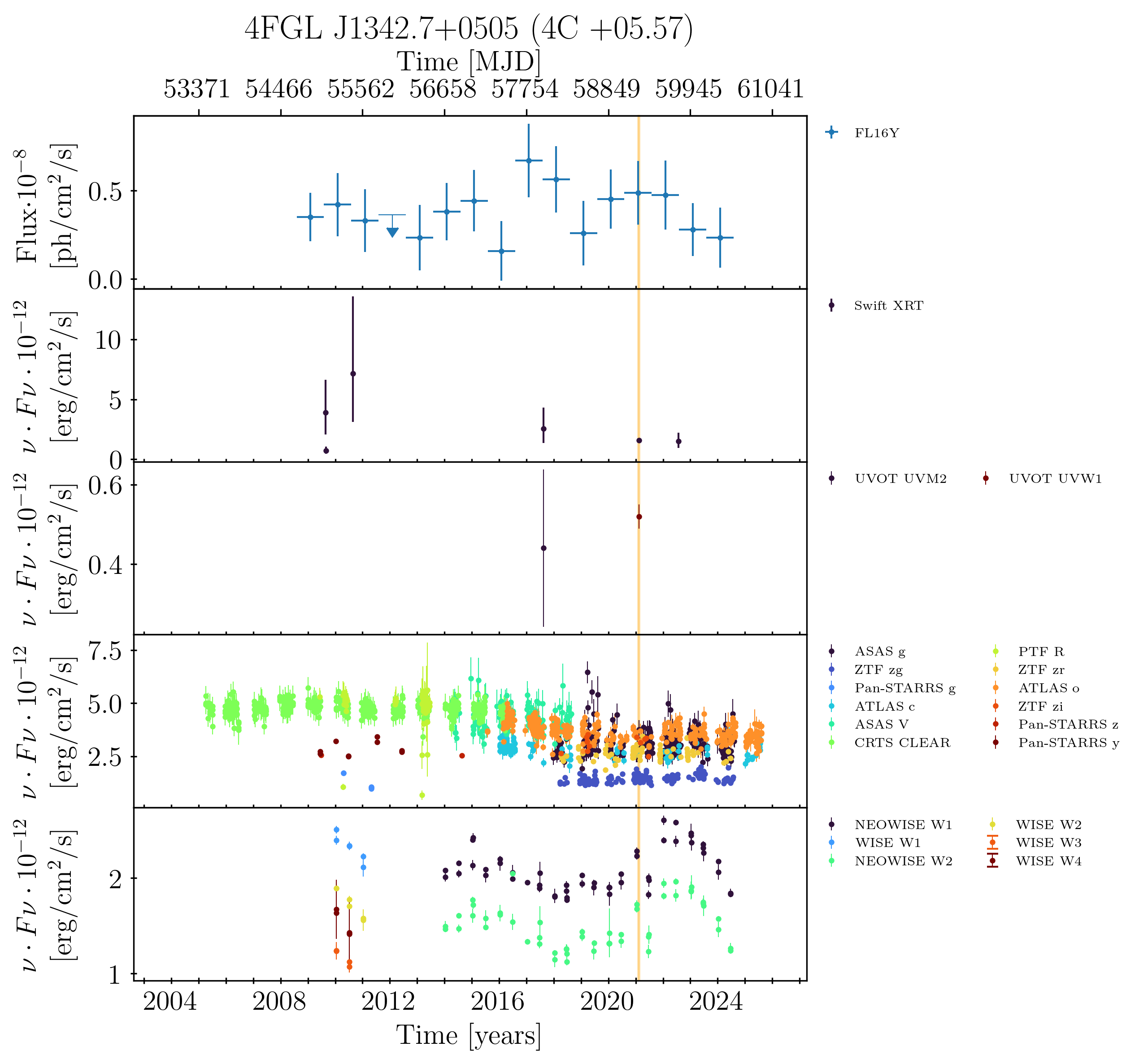}
    \caption{Multifrequency light curves of \J1342. The panels display from top to bottom: \textit{Fermi}-LAT gamma-ray light curves integrated over a 12-months time bin taken from the FL16Y source list; X-ray data from \textit{Swift-}XRT; UV data from \textit{Swift-}UVOT; optical data from different facilities; IR data from WISE/NEOWISE. The orange stripe highlights the arrival time of the neutrino candidate associated with this source (IC210210A).}
    \label{fig:LC_J1342p0504}
\end{figure*}

The following plots display the collected multifrequency data (see Section \ref{sec: observational data}) for the primary sample defined in Section \ref{sec: sample}. The sources show modest variability with flux changes generally in the range of two to three times the baseline flux. Two sources show variability close to the neutrino arrival time, which are \J1258 (see Figure \ref{fig:LC_J1258m0447}) and \J1342 (see Figure \ref{fig:LC_J1342p0504}). For the latter one, the flux change is only visible in the IR regime while the optical flux seems to remain flat. This could be explained by different emission regions as visible in Figure \ref{fig:lepto-hadronic_solutions}, where the optical flux is dominated by the host galaxy, while the IR data are attributed to jet emission.
\begin{figure*}[h]
    \centering
    \includegraphics[width=1\linewidth]{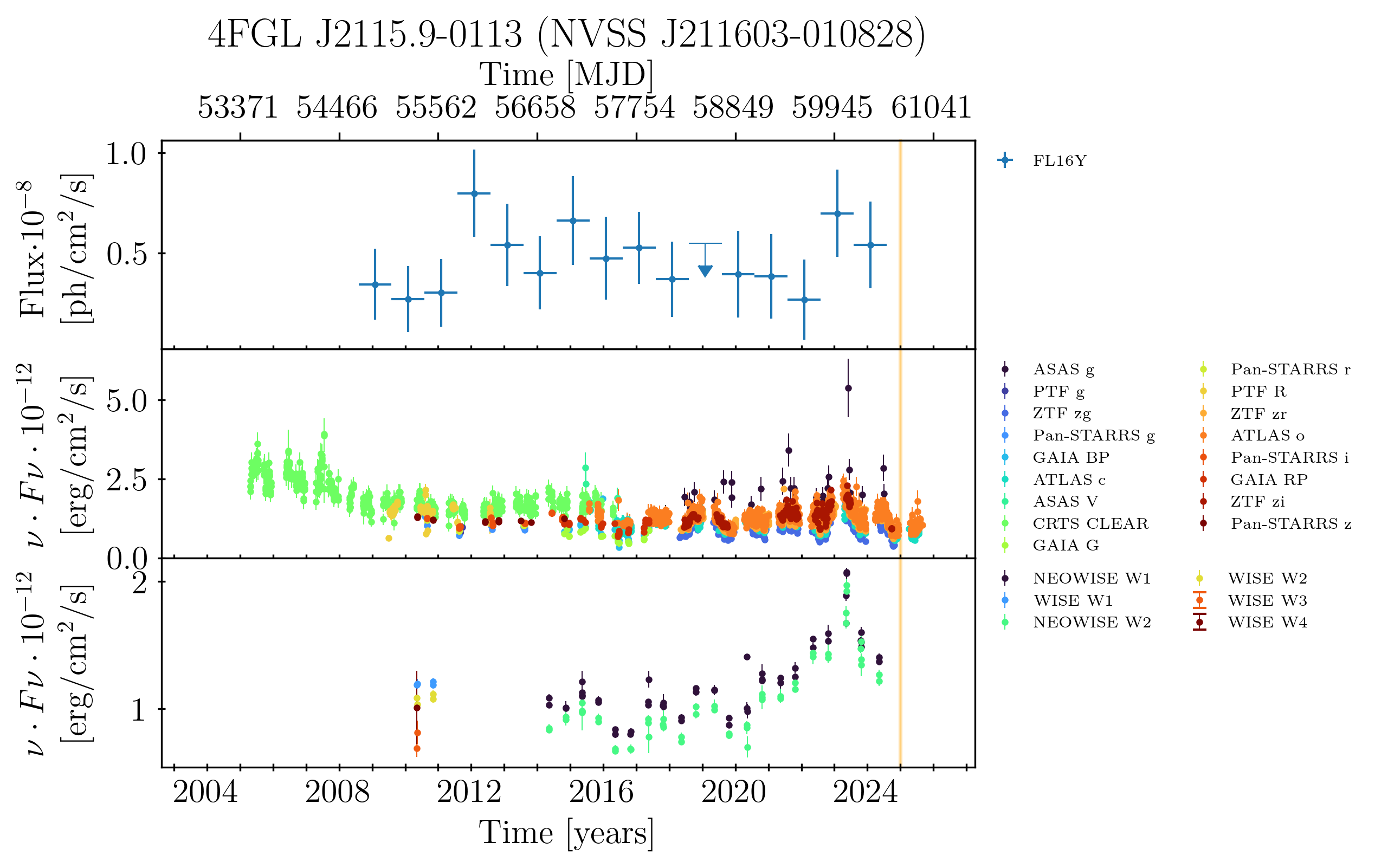}
    \caption{Multifrequency light curves of \J2115. The panels display from top to bottom: \textit{Fermi}-LAT gamma-ray light curves integrated over a 12-months time bin taken from the FL16Y source list; optical data from different facilities; IR data from WISE/NEOWISE. The orange stripe highlights the arrival time of the neutrino candidate associated with this source (IC250101A).}
    \label{fig:LC_J2116m0108}
\end{figure*}
\begin{figure*}[h]
    \centering
    \includegraphics[width=1\linewidth]{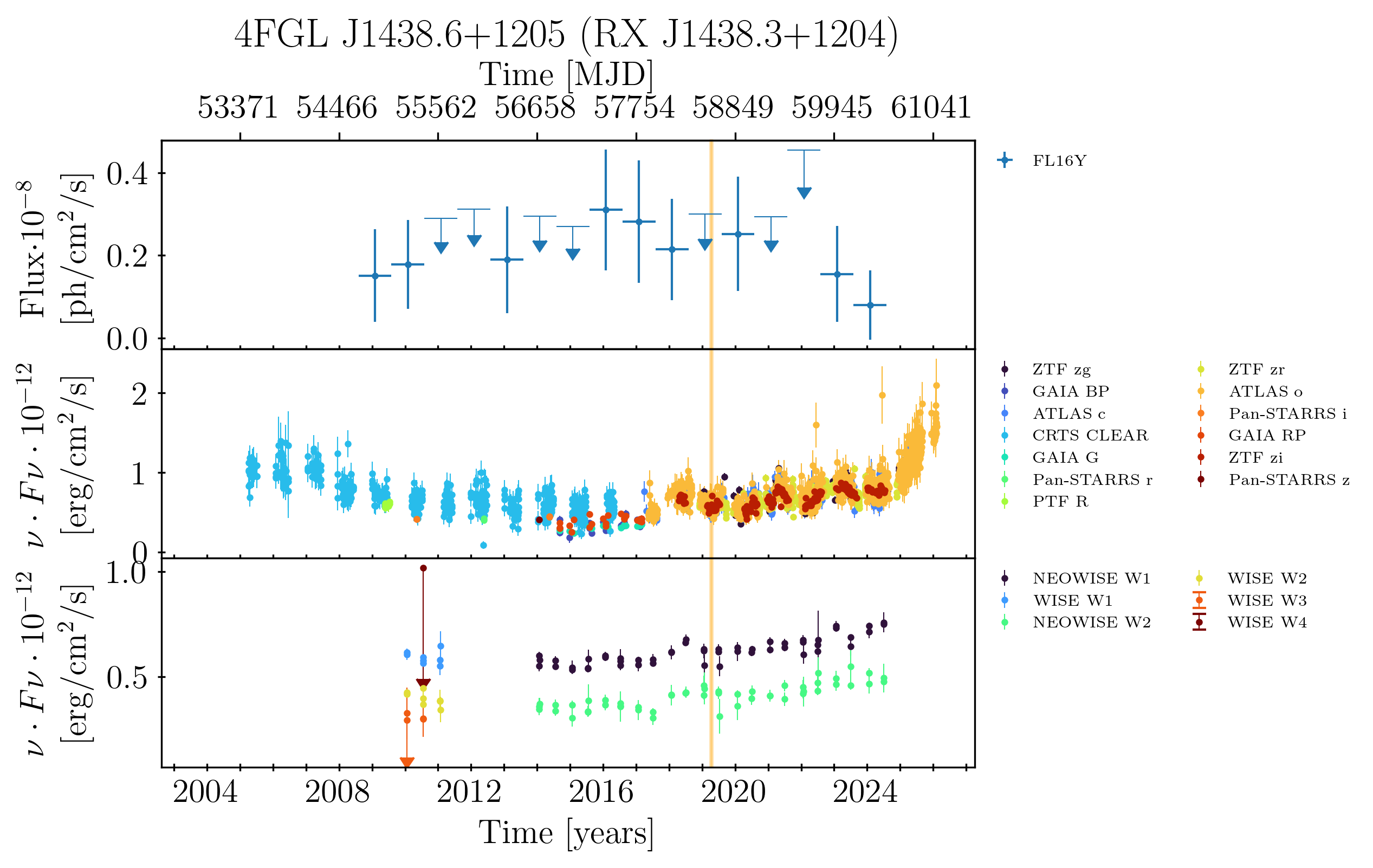}
    \caption{Multifrequency light curves of \J1438.6. The panels display from top to bottom: \textit{Fermi}-LAT gamma-ray light curves integrated over a 12-months time bin taken from the FL16Y source list; optical data from different facilities; IR data from WISE/NEOWISE. The orange stripe highlights the arrival time of the neutrino candidate associated with this source (IC190413A).}
    \label{fig:LC_J1438p1204}
\end{figure*}
\begin{figure*}[h]
    \centering
    \includegraphics[width=1\linewidth]{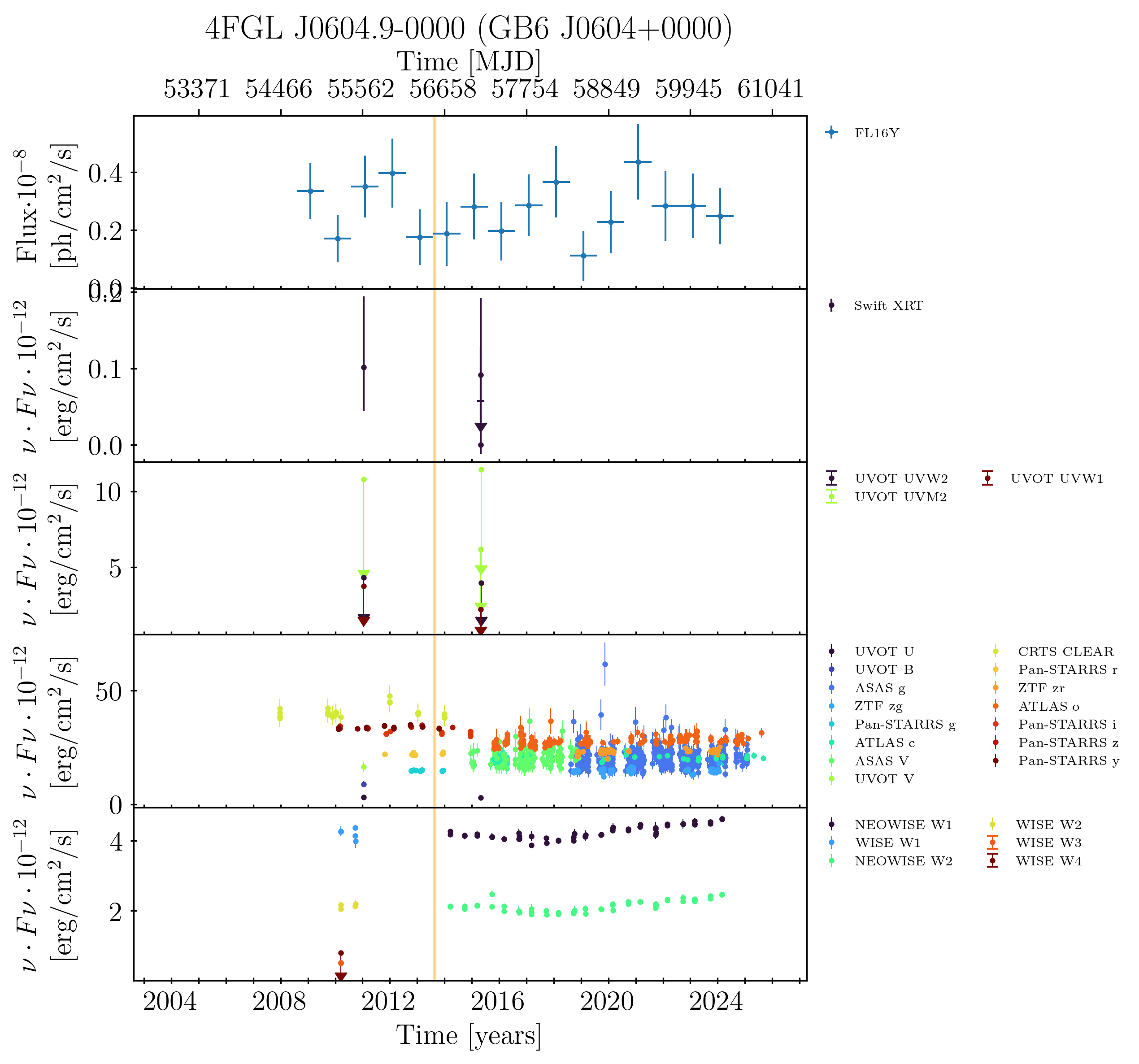}
    \caption{Multifrequency light curves of \J0604. The panels display from top to bottom: \textit{Fermi}-LAT gamma-ray light curves integrated over a 12-months time bin taken from the FL16Y source list; X-ray data from \textit{Swift-}XRT; UV data from \textit{Swift-}UVOT; optical data from different facilities; IR data from WISE/NEOWISE. The orange stripe highlights the arrival time of the neutrino candidate associated with this source (IC130822A).}
    \label{fig:LC_J0604p0000}
\end{figure*}
\begin{figure*}[h]
    \centering
    \includegraphics[width=1\linewidth]{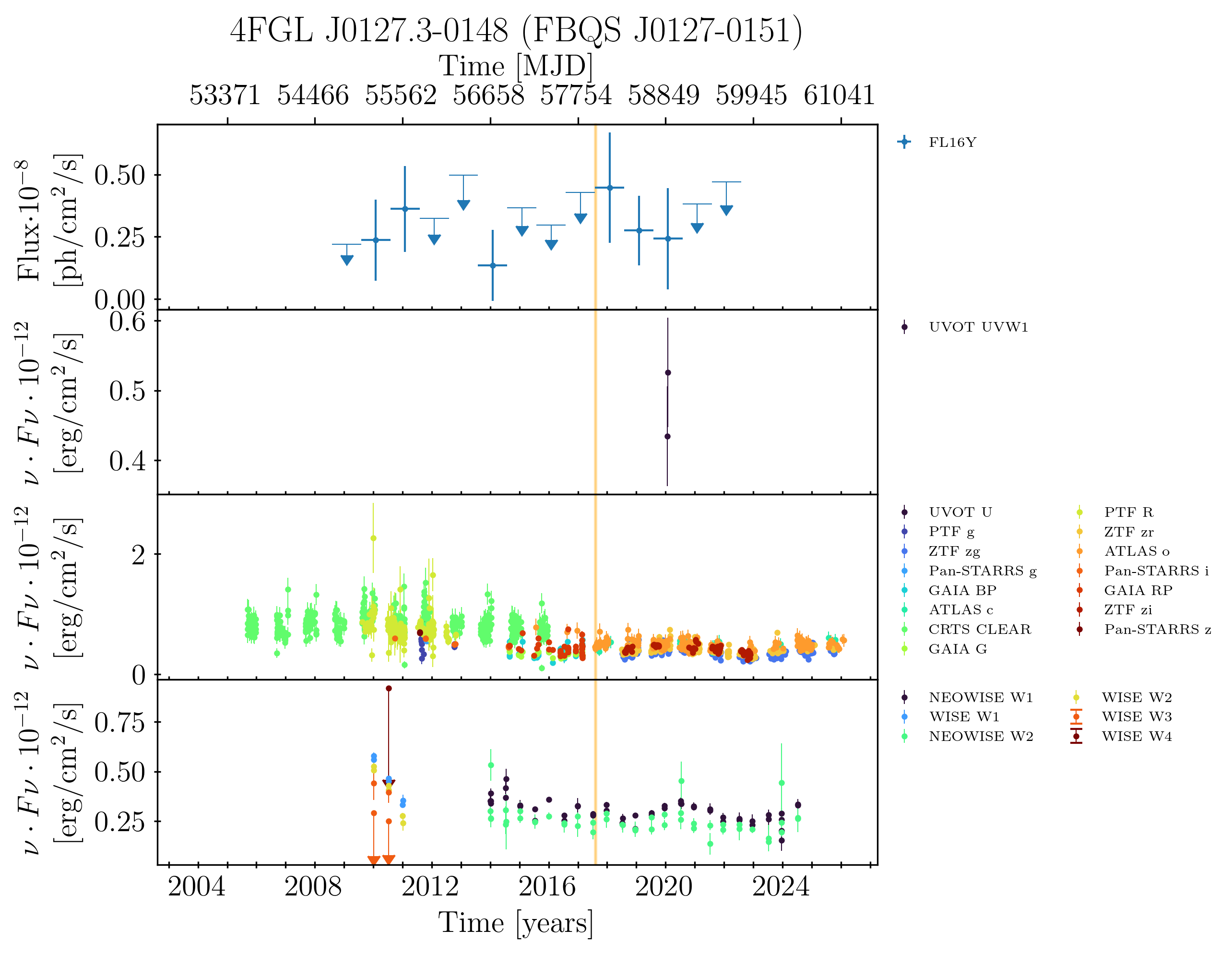}
    \caption{Multifrequency light curves of \J0127. The panels display from top to bottom: \textit{Fermi}-LAT gamma-ray light curves integrated over a 12-months time bin taken from 4FGL-DR4 catalog; UV data from \textit{Swift-}UVOT; optical data from different facilities; IR data from WISE/NEOWISE. The orange stripe highlights the arrival time of the neutrino candidate associated with this source (IC170809A).}
    \label{fig:LC_J0127m0151}
\end{figure*}
\end{appendix}
\clearpage
\twocolumn

\end{document}